\documentclass[longauth]{aa} 

\usepackage{graphicx}
\usepackage{color}
\usepackage{soul}
\usepackage{txfonts}
\usepackage{float}
\usepackage[T1]{fontenc}
\usepackage{caption}
\usepackage{rotating}
\usepackage{booktabs}
\usepackage{afterpage}
\usepackage[para]{threeparttable}
\usepackage{subcaption}
\usepackage{caption}
\usepackage[]{hyperref}
   \newcommand{\asec}{^{\prime\prime}}
   
   \newcommand{\beam}{$\theta_{\mbox{\scriptsize maj}}\times\theta_{\mbox{\scriptsize min}}$}

\DeclareCaptionFormat{cont}{#1 (cont.)#2#3\par}

\begin{document}

   \title{ALMA survey of Class II protoplanetary disks in Corona Australis: a young region with low disk masses\thanks{Based on observations made with ESO Telescopes at the La Silla
Paranal Observatory under programme ID 299.C-5048 and 0101.C-0893}}
   \subtitle{}

  \author{P. Cazzoletti\inst{1}
          \and C. F. Manara\inst{2}
           \and Hauyu Baobab Liu\inst{2,3}
           \and E. F. van Dishoeck\inst{1,4}
          \and S. Facchini\inst{2}
          \and J. M. Alcalà\inst{5}
          \and M. Ansdell \inst{6,7}
          \and L. Testi\inst{2}
          \and J. P. Williams\inst{8}
          \and C. Carrasco-Gonz\'alez\inst{9}
          \and R. Dong\inst{10}
          \and J. Forbrich \inst{11,12}
          \and M. Fukagawa \inst{13}
          \and R. Galván-Madrid\inst{9}
          \and N. Hirano\inst{3}
          \and M. Hogerheijde\inst{4,14}
          \and Y. Hasegawa\inst{15}
          \and T. Muto\inst{16}
          \and P. Pinilla\inst{17}
          \and M. Takami\inst{3}
          \and M. Tamura\inst{13,18,19}
          \and M. Tazzari\inst{20}
          \and J. P. Wisniewski\inst{21}
          }

   \institute{
      Max-Planck-Institute for Extraterrestrial Physics (MPE), Giessenbachstr. 1, 85748, Garching, Germany \\
       \email{pcazzoletti@mpe.mpg.de}
      \and
      European Southern Observatory (ESO), Karl-Schwarzschild-Str. 2, D-85748 Garching, Germany 
  	\and
	Academia Sinica Institute of Astronomy and Astrophysics, Roosevelt Rd, Taipei 10617, Taiwan
	\and
	Leiden Observatory, Leiden University, Niels Bohrweg 2, 2333 CA Leiden, The Netherlands
	\and
	INAF-Osservatorio Astronomico di Capodimonte, via Moiariello 16, 80131 Napoli, Italy 
	\and
	Center for Integrative Planetary Science, University of California at Berkeley, Berkeley, CA 94720, USA
         \and
         Department of Astronomy, University of California at Berkeley, Berkeley, CA 94720, USA
         \and
         Institute for Astronomy, University of Hawai`i at M\={a}noa, Honolulu, HI 96822, USA
         \and
	Instituto de Radioastronom\'{\i}a y Astrof\'{\i}sica (IRyA-UNAM), Universidad Nacional Autónoma de México, Campus Morelia, Apartado Postal 3-72, 58090 Morelia, Michoacán, Mexico
	\and
	Department of Physics \& Astronomy, University of Victoria, Victoria, BC, V8P 1A1, Canada
	\and
	Centre for Astrophysics Research, University of Hertfordshire, College Lane, Hatfield AL10 9AB, UK
	\and
	Harvard-Smithsonian Center for Astrophysics, 60 Garden St, Cambridge, MA 02138, USA
	\and
	National Astronomical Observatory of Japan, 2-21-1 Osawa, Mitaka, Tokyo 181-8588 Japan
	\and
	Anton Pannekoek Institute for Astronomy, University of Amsterdam, PO Box 94249, 1090 GE, Amsterdam, the Netherlands
	\and
	Jet Propulsion Laboratory, California Institute of Technology, Pasadena, CA 91109, USA
	\and
	 Division of Liberal Arts, Kogakuin University, 1-24-2 Nishi-Shinjuku, Shinjuku-ku, Tokyo 163-8677, Japan
	 \and
	 Department of Astronomy/Steward Observatory, The University of Arizona, 933 North Cherry Avenue, Tucson, AZ 85721, USA
	 \and
	 Department of Astronomy, The University of Tokyo, 7-3-1, Hongo, Bunkyo-ku, Tokyo 113-0033, Japan
	 \and
	 Astrobiology Center of NINS, 2-21-1, Osawa, Mitaka, Tokyo 181-8588, Japan
	\and
	Institute of Astronomy, University of Cambridge, Madingley Road, Cambridge CB3 0HA, UK
	\and
	Homer L. Dodge Department of Physics and Astronomy, University of Oklahoma, 440 W. Brooks Street, Norman, OK 73019, USA
             }

   \date{Received December XX, 2017; accepted September XX, 2017}

  \abstract
    {In recent years,  the disk populations in a number of young star-forming regions have been surveyed with the Atacama Large Millimeter/submillimeter Array (ALMA). Understanding the disk properties and their correlation with the properties of the central star is critical to understand planet formation. In particular, a decrease of the average measured disk dust mass with the age of the region has been observed, consistent with grain growth and disk dissipation.} 
   {We want to compare the general properties of disks and their host stars in the nearby ($d=160\,\rm pc$) Corona Australis (CrA) star forming region to those of the disks and stars in other regions.} 
   {We conducted high-sensitivity continuum ALMA observations of 43 Class II young stellar objects in CrA at 1.3 mm (230 GHz). The  typical spatial resolution is $\sim 0.3\asec$. The continuum fluxes are used to estimate the dust masses of the disks, and a survival analysis is performed to estimate the average dust mass. We also obtained new VLT/X-Shooter spectra for 12 of the objects in our sample for which spectral type information was missing.
   }
   {24 disks are detected, and stringent limits have been put on the average dust mass of the non-detections. Taking into account the upper limits, the average disk mass in CrA is $6\pm3\,\rm M_\oplus$. This value is significantly lower than that of disks in other young (1-3 Myr) star forming regions (Lupus, Taurus, Chamaeleon I, and Ophiuchus) and appears to be consistent with the average disk mass of the 5-10 Myr old Upper Sco. The position of the stars in our sample on the Herzsprung-Russel diagram, however, seems to confirm that that CrA has age similar to Lupus. Neither external photoevaporation nor a lower than usual stellar mass distribution can explain the low disk masses. On the other hand, a low-mass disk population could be explained if the disks are small, which could happen if the parent cloud has a low temperature or intrinsic angular momentum, or if the the angular momentum of the cloud is removed by some physical mechanism such as magnetic braking. Even in detected disks, none show clear substructures or cavities.} 
   {Our results suggest that in order to fully explain and understand the dust mass distribution of protoplanetary disks and their evolution, it may also be necessary to take into consideration the initial conditions of star and disk formation process. These conditions at the very beginning may potentially vary from region to region, and could play a crucial role in planet formation and evolution.}

   \keywords{protoplanetary disks --- submillimeter: ISM --- planets and satellites: formation --- stars: pre-main sequence --- 
              stars: variables: T Tauri, Herbig Ae/Be --- stars: formationt }

\titlerunning{ALMA band 6 survey for Class II YSOs in CrA}

   \maketitle

\section{Introduction}\label{sec:introduction}
Planets form in protoplanetary disks around young stars, and the way these disks evolve also impacts what kind of planetary system will be formed \citep{2016JGRE..121.1962M}. The evolution of the disk mass with time is one of the key ingredients of planetary synthesis models \citep{2014prpl.conf..691B}.
For a long time infrared telescopes (e.g., \textit{Spitzer}) have shown how the inner regions of disks dissipate on a timescale of $\sim$3-5 Myr \citep{2001ApJ...553L.153H, 2007ApJ...662.1067H,2010A&A...510A..72F,2013MNRAS.434..806B}. 

\begin{figure*}
\centering
\includegraphics[width=\textwidth]{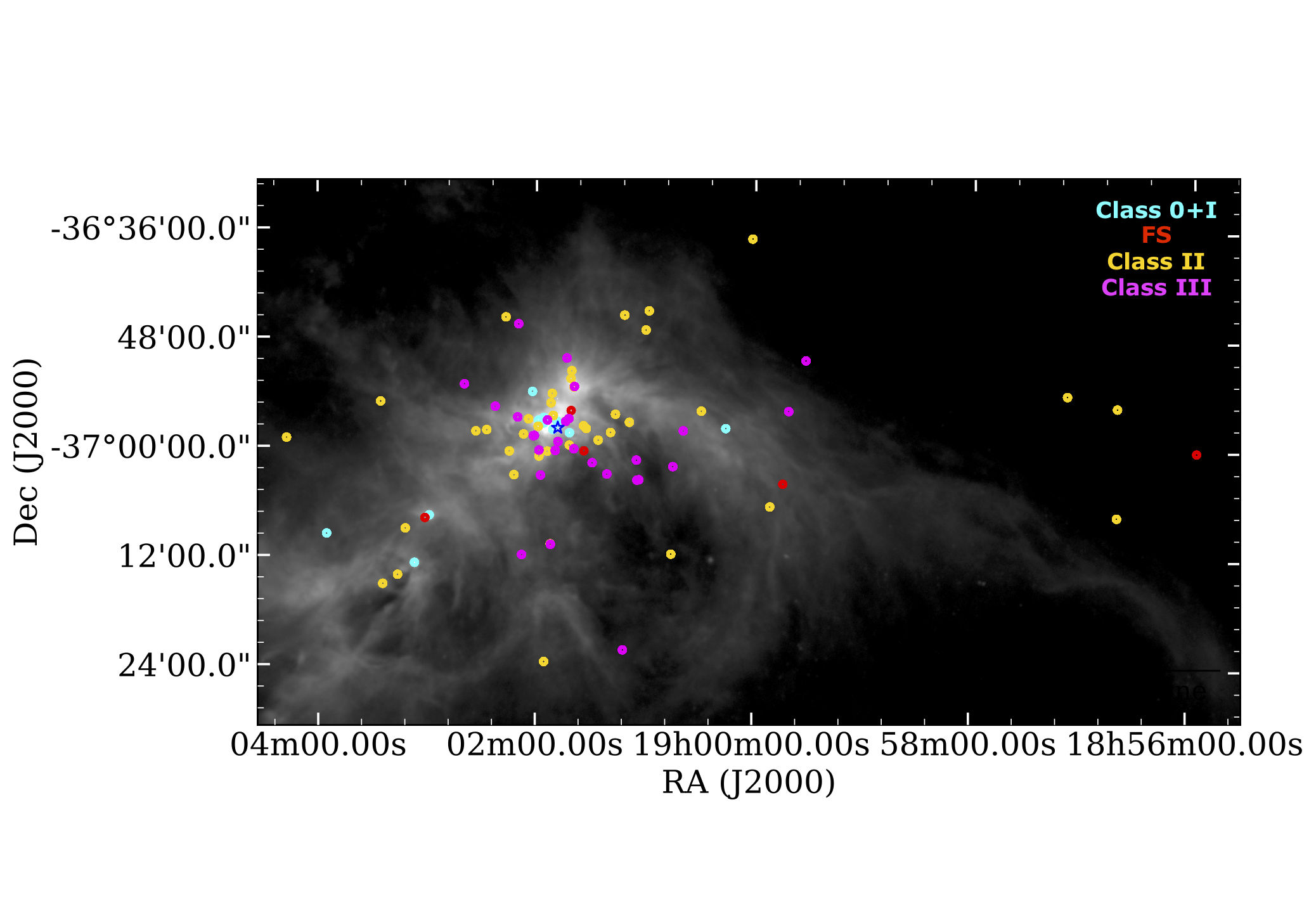}
\caption{Spatial distribution of the CrA sources from the \citet{peterson} catalogue on top of the \textit{Herschel} $250\,\rm \mu m$ map of the Corona Australis molecular Cloud. The different colours represent the classification of the YSOs. The blue star indicates the position of R CrA}
\label{fig:map_classes}
\end{figure*} 

Only recently, however, we have been able to measure the bulk disk mass for statistically significant samples of disks, thanks to the high sensitivity of the Atacama Large Millimeter/submillimeter Array (ALMA). Pre-ALMA surveys of disk masses were restricted to the northern hemisphere Taurus, Ophiuchus and Orion Nebula Cluster regions \citep{2005ApJ...631.1134A, 2009ApJ...700.1502A,andrews2013, 2008ApJ...683..304E, 2010ApJ...725..430M}. In the first years of operations of ALMA this has changed dramatically: hundreds of disks have been surveyed to determine the disk population in the $\sim$1-3 Myr old Lupus, Chamaeleon~I, Orion Nebula Cluster, Ophiuchus, IC348 and Taurus regions \citep{lupus,2016ApJ...831..125P,2018ApJ...860...77E,2019MNRAS.482..698C,2018MNRAS.478.3674R,2018ApJ...869...17L}, in the $\sim$3-5 Myr old $\sigma$-Orionis region \citep{2017AJ....153..240A}, and in the older $\sim$5-10 Myr Upper Scorpius association \citep{2016ApJ...827..142B}. These surveys have shown that the typical mass of protoplanetary disks decreases with the age of the region, in line with  the observations that the inner regions of disks are dissipated within $\sim$ 3-5 Myr, similar to the dissipation time scale measured in the infrared. A positive correlation between disk and stellar mass was also found, and a steepening of its slope with time was identified \citep{lupus,2017AJ....153..240A,2016ApJ...831..125P}.  This is consistent with the result that massive planets form and are found preferentially around more massive stars \cite[e.g.][]{2013A&A...549A.109B, 2011A&A...526A..63A}. Finally, the steepening of the relation with time is explained with more efficient radial drift around low mass stars \citep{2016ApJ...831..125P}, and it suggests that a significant portion of the planet formation process, especially around low mass stars, must happen in the first $\sim$1-2 Myr, when enough material to form planets is still available in disks \citep{2016A&A...593A.111T,2018A&A...618L...3M}. Studying the evolution  of the $\rm M_{\rm disk}-M_\star$ relation in as many different environments as possible is therefore critical for understanding how the planet formation process is affected by the mass of the central stars.

We present here a survey of the Class II disks in the Corona Australis star forming region (CrA). Located at an average distance of about 154 pc \citep{2018A&A...616A...1G,2018ApJ...867..151D}, the CrA molecular cloud complex is one of the nearest star-forming regions \citep[see review in][]{2008hsf2.book..735N}. It has been the target of many infrared surveys, the most recent being the Gould Belt (GB) \textit{Spitzer} Legacy program presented in \citet{peterson}. At the center of the CrA region is located the \textit{Coronet} cluster, which is a region of young embedded objects in the vicinity of R CrA \citep[Herbig Ae star, ][]{2000A&AS..146..323N}, on which many of the previous studies have focused. All studies agree in assigning to the \textit{Coronet} an age $<3\,\rm Myr$ \citep[e.g.][]{2009PASP..121..350M,2011ApJ...736..137S}. However, there are also some indications of a more evolved population \citep[e.g.][]{2000A&AS..146..323N,peterson, 2011ApJ...736..137S}. A deep, sub-mm wavelength survey of the disk population in the region can help to further understand the formation and evolutionary history of CrA.

We therefore use ALMA to conduct a high-sensitivity millimeter wavelength survey of all the known Class II sources in CrA and compare the results with other regions surveyed to-date. In Sec. \ref{sec:sample} the sample is described, while the ALMA  observations are detailed in Sec. \ref{sec:obs}. We also describe there new VLT/X-Shooter observations to determine the stellar charachteristics. The continuum millimeter measurements, their conversion to dust masses and a comparison with other star-forming regions is presented in Sec. \ref{sec:results}. Our findings are interpreted in the context of disk evolution in Sec. \ref{sec:discussion}. Finally, the work is summarized in Sec. \ref{sec:conclusion}.

\begin{table*}[h!]
\center
\caption{Stellar properties of the central sources of the disks in the sample. The RA and DEC in J2000 are from the \textit{Spitzer} data presented in \citet{peterson}}\label{tab:stellarprop}
\begin{threeparttable}
\begin{tabular}{llcccccccc}
\toprule
2MASS ID & Name & RA & DEC & SpT & Ref. \\
\midrule
J18563974-3707205 &    CrA-1  	            &	 18:56:39.76    &	 -37:07:20.8 	&	 M6  	&	 1 \\
J18595094-3706313 &    CrA-4  	            &	 18:59:50.95    &	 -37:06:31.6 	&	 M8   	&	 2 \\
J19002906-3656036 &    CrA-6  	            &	 19:00:29.07    &	 -36:56:03.8 	&	 M4   	&	 3\\
J19004530-3711480 &    CrA-8  	            &	 19:00:45.31    &	 -37:11:48.2 	&	 M8.5 	&	 4 \\
J19005804-3645048 &    CrA-9  	            &	 19:00:58.05    &	 -36:45:05.0 	&	 M1   	&	 3 \\
J19005974-3647109 &    CrA-10 	            &	 19:00:59.75    &	 -36:47:11.2 	&	 M4   	&	 5 \\
J19011629-3656282 &    CrA-12 	            &	 19:01:16.29    &	 -36:56:28.3 	&	 M5   	&	 6 \\
J19011893-3658282 &    CrA-13 	            &	 19:01:18.95    &	 -36:58:28.4 	&	 M2   	&	 7 \\
J19013232-3658030 &    CrA-15 	            &	 19:01:32.31    &	 -36:58:03.0 	&	 M3.5 	&	 7 \\
J19013385-3657448 &    CrA-16 	            &	 19:01:33.85    &	 -36:57:44.9 	&	 M2.5 	&	 7 \\
J19014041-3651422 &    CrA-18 	            &	 19:01:40.41    &	 -36:51:42.3 	&	 M1.5 	&	 7 \\
J19015112-3654122 &    CrA-21 	            &	 19:01:51.12    &	 -36:54:12.4 	&	 M2 	&	 8 \\
J19015180-3710478 &    CrA-22 	            &	 19:01:51.86    &	 -37:10:44.7 	&	 M4.5 	&	 1 \\
J19015374-3700339 &    CrA-23 	            &	 19:01:53.75    &	 -37:00:33.9 	&	 M7.5 	&	 7 \\
J19020682-3658411 &    CrA-26 	            &	 19:02:06.80    &	 -36:58:41.0 	&	 M7 	&	 1 \\
J19021201-3703093 &    CrA-28 	            &	 19:02:12.00    &	 -37:03:09.4 	&	 M4.5 	&	 5 \\
J19021464-3700328 &    CrA-29 	            &	 19:02:14.63    &	 -37:00:32.9 	&	 ... 	&	 ... \\
J19022708-3658132 &    CrA-30 	            &	 19:02:27.07    &	 -36:58:13.1 	&	 M0.5 	&	 5 \\
J19023308-3658212 &    CrA-31 	            &	 19:02:33.07    &	 -36:58:21.2 	&	 M3.5 	&	 1 \\
J19031185-3709020 &    CrA-35 	            &	 19:03:11.84    &	 -37:09:02.1 	&	 M5 	&	 3 \\
J19032429-3715076 &    CrA-36 	            &	 19:03:24.29    &	 -37:15:07.7 	&	 M5 	&	 1 \\
J19012576-3659191 &    CrA-40 	            &	 19:01:25.75    &	 -36:59:19.1 	&	 M4.5	&	 1 \\
J19014164-3659528 &    CrA-41 	            &	 19:01:41.62    &	 -36:59:52.7 	&	 M2 	&	 9 \\
J19015037-3656390 &    CrA-42 	            &	 19:01:50.48    &	 -36:56:38.4 	&	 ... 	&	 ... \\
J19031609-3714080 &    CrA-45 E 	        &	 19:03:16.09 	&	 -37:14:08.2 	&	 M3.5 	&	 1 \\
J19031609-3714080 &    CrA-45 W 	        &	 19:03:16.09 	&	 -37:14:08.2 	&	 M3.5 	&	 1 \\
J18564024-3655203 &    CrA-47 	            &	 18:56:40.28    &	 -36:55:20.8 	&	 M6 	&	 1 \\
J18570785-3654041 &    CrA-48 	            &	 18:57:07.86    &	 -36:54:04.4 	&	 M5	&	 1 \\
J19000157-3637054 &    CrA-52 	            &	 19:00:01.58    &	 -36:37:06.2 	&	 M1 	&	 10 \\
J19011149-3645337 &    CrA-53 	            &	 19:01:11.49    &	 -36:45:33.8 	&	 M5 	&	 1 \\
J19013912-3653292 &    CrA-54 	            &	 19:01:39.15    &	 -36:53:29.4 	&	 K7 	&	 9 \\
J19015523-3723407 &    CrA-55 	            &	 19:01:55.23    &	 -37:23:41.0 	&	 K5 	&	 11 \\
J19021667-3645493 &    CrA-56 	            &	 19:02:16.66    &	 -36:45:49.4 	&	 M4 	&	 4 \\
J19032547-3655051 &    CrA-57 	            &	 19:03:25.48    &	 -36:55:05.3 	&	 M4.5 	&	 1 \\
J19010860-3657200 &    SCrA N               &	 19:01:08.62 	&	 -36:57:20 	    &	 K3 &	 6 \\
J19010860-3657200 &    SCrA S               &	 19:01:08.62 	&	 -36:57:20 	    &	 M0 &	 6 \\
J19015878-3657498 &    TCrA         	    &	 19:01:58.78 	&	 -36:57:49 	    &	  F0 	&	 6 \\
J19014081-3652337 &    TYCrA  	            &	 19:01:40.83 	&	 -36:52:33.88 	&	 B9	&	 6 \\
J19041725-3659030 &    Halpha15     	    &	 19:04:17.25 	&	 -36:59:03.0 	&	 M4 	&	 12\\
J19025464-3646191 &    ISO-CrA-177  	    &	 19:02:54.65 	&	 -36:46:19.1 	&	 M4.5 	&	 4 \\
...               &    G09-CrA-9    	    &	 19:01:58.34 	&	 -37:01:06.0 	&	 ... 	&	 ... \\
J19015173-3655143 &    Haas17       	    &	 19:01:51.74 	&	 -36:55:14.2 	&	 ... 	&	 ... \\
J19020410-3657013 &    IRS10        	        &	 19:02:04.09 	&	 -36:57:01.2 	&	 ... 	&	 ... \\
\bottomrule
\end{tabular}
\label{default}
\begin{tablenotes}
 \small
 \item \textbf{References.} (1) This work,  (2) \citet{2004A&A...424..213B}, (3) \citet{2012ApJ...749...79R}, (4) \citet{2005A&A...444..175L}, (5) \citet{2011ApJ...736..137S}, (6) \citet{2007A&A...475..959F}, (7) \citet{2008ApJ...687.1145S}, (8) \citet{2011ApJ...732...24C}, (9) \citet{2009PASP..121..350M}, (10) \citet{1997AJ....114.1544W}, (11) \citet{2014ApJ...786...97H}, (12) \citet{1998ASPC..154.1755P}
\end{tablenotes}
\end{threeparttable}
\end{table*}

\section{Sample selection}\label{sec:sample}
\citet{peterson} present in their work a comprehensive catalogue of known Young Stellar Objects (YSOs) in the CrA star forming region selected based on \textit{Spitzer}, 2MASS, \textit{ROSAT}, and \textit{Chandra} data. In addition to the these, they also added more YSOs from the literature. Their final catalogue includes a  total of 116 YSOs, 14 of which are classified as Class I, 5 as Flat Spectrum (FS), 43 as Class II and 54 as Class III. The Infrared Class was determined  by calculating the spectral slope $\alpha$ over the widest possible range of IR wavelengths as follows:
\begin{equation}
\alpha=\frac{\Delta\log{(\lambda F_\lambda)}}{\Delta\log{\lambda}},
\end{equation}
where $\lambda$ is the wavelength and $F_\lambda$ the flux at $\lambda$. Sources with $\alpha\geq0.3$ are classified as Class I; FS have $-0.3\leq\alpha<0.3$; Class II have $-1.6\leq\alpha<-0.3$; sources with $\alpha<-1.6$ are Class III \citep{2009ApJS..181..321E,peterson}. Fig. \ref{fig:map_classes} shows the spatial distribution of the sources and their classification on top of the \textit{Herschel} $250\,\rm \mu m$ map of the molecular Cloud.

\begin{figure}[]
\includegraphics[width=\columnwidth]{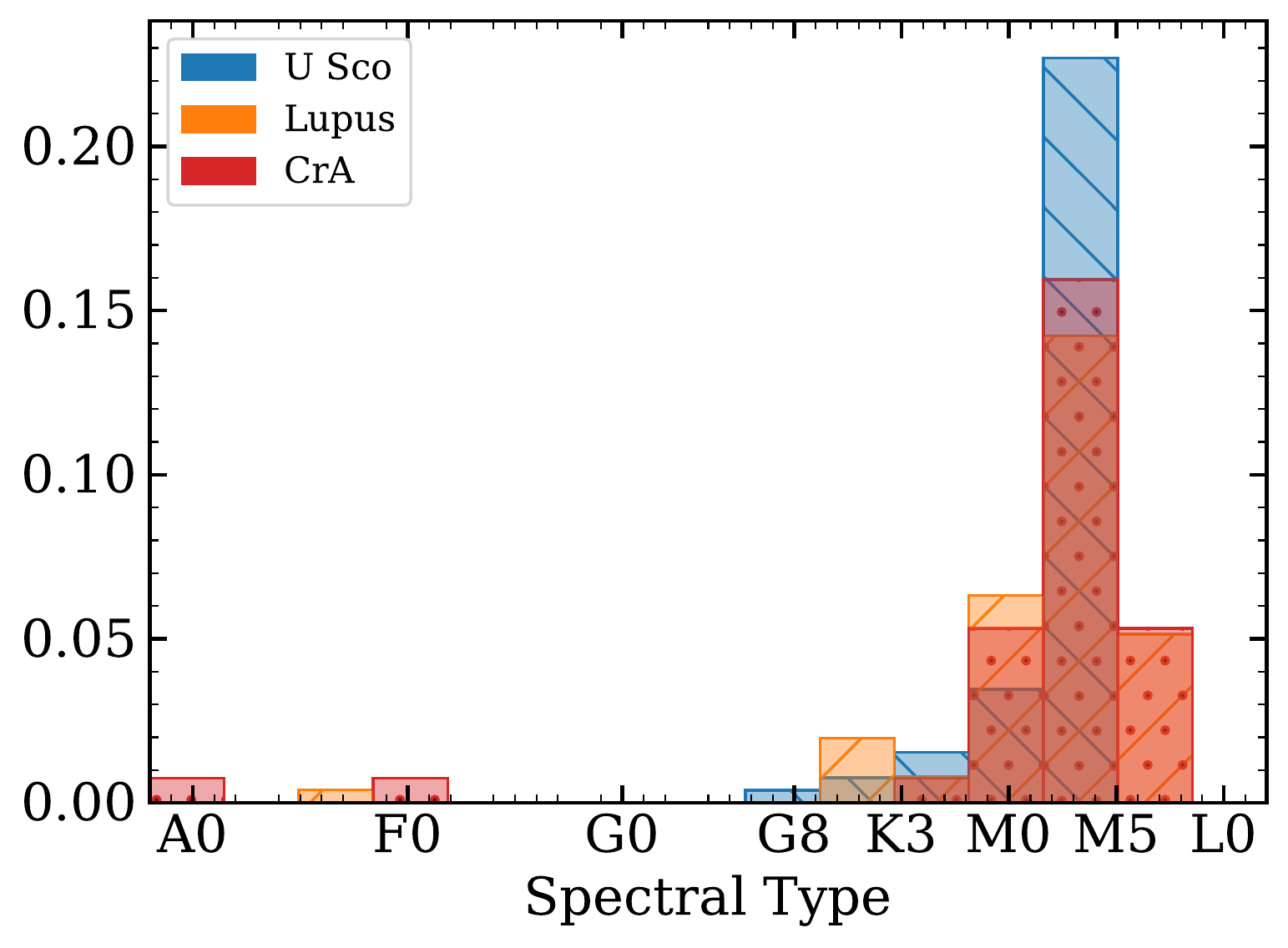}
\caption{
Distribution of the spectral types of the stars in CrA (Red) compared to that of Lupus (Orange) and Upper Sco (Blue).
}
\label{fig:spt_dist}
\end{figure}

Our sample includes all the Class II sources from the \citet{peterson} catalogue. Two of them (CrA-49 and CrA-51) were later identified as background, evolved stars based on parallax measurements with Gaia \citep{2016A&A...595A...1G, 2018arXiv180409366L, 2018A&A...616A...9L,2018A&A...616A...1G} and on our VLT/X-Shooter spectra (see Sec. \ref{sec:data_xshooter}). 
We then checked our sample against the more recently published survey by \citet{dunham} in which the \textit{Spitzer} data are re-analysed and the spectral slopes re-calculated. We find broad agreement between the classification in \citet{peterson} and \citet{dunham}, except for a few very marginal cases at the boundaries of classes.

Our final sample contains 41 targets, two of which are clearly resolved binaries (S CrA and CrA-45). Of the 43 targeted disks, 24 are detected with ALMA. The spectral type (SpT) was known for only 26 of the stars from the literature. We obtained VLT/X-Shooter spectra for 11 of the remaining targets, and derived their properties as explained in Sec. \ref{sec:stellarprop}.

The basic stellar properties for the CrA sample are given in Table \ref{tab:stellarprop}, the distribution of SpTs is shown in Fig. \ref{fig:spt_dist}, while the millimeter observations, flux densities, and calculated disk masses are presented in Table \ref{tab:millimeter}.

\section{Observations}\label{sec:obs}
\subsection{ALMA observations}
We have carried out three executions of observations at 1.3 mm towards 43 Class\,II YSOs in the Corona Australis molecular cloud, using ALMA (2015.1.01058.S, PI: H. B. Liu)..
Each one of the 43 target sources were integrated for approximately 1 minute in each epoch.
The spectral setup consists of six spectral windows, of which the (central frequency [GHz], total bandwidth [MHz], and frequency channel width [kHz]) are (216.797, 1875, 488), (219.552, 59, 61), (219.941, 59, 61), (220.390, 117, 61), (230.531, 117, 31), (231.484, 1875, 488), respectively.
 Additional observational details are summarized in Table \ref{tab:observations}. $^{12}$CO (2-1), $^{13}$CO (2-1) and C$^{18}$O (2-1) transitions were also targeted with our spectral setup, but no clear detection was found because of strong foreground contamination. SO (6-5) and SiO (5-4) lines were also covered and not detected.

The data were manually calibrated using the CASA v5.1.1 software package \citep{2007ASPC..376..127M}	.
The gain calibrator for the first epoch of observations was faint.
To yield reasonably high signal-to-noise (S/N) ratios when deriving the gain phase solutions,  the phase offsets among spectral windows  were first solved using the passband calibration scan.
After applying the phase offsets solution, the gain phase solution was then derived  by combining all spectral windows.
The calibration of the other two epochs of observations followed the standard procedure of ALMA quality assurance (i.e., QA2).
The bootstrapped flux values of the calibrator quasar J1924-2914 were consistent with the SMA Calibrator list\footnote{http://sma1.sma.hawaii.edu/callist/callist.html} \citep{2007ASPC..375..234G} to $\sim$10\%.
After calibration, we fit the continuum baseline and subtract it from the spectral line data, using the CASA task {\tt uvcontsub}.

\begin{figure*}
\centering
\includegraphics[width=0.95\textwidth]{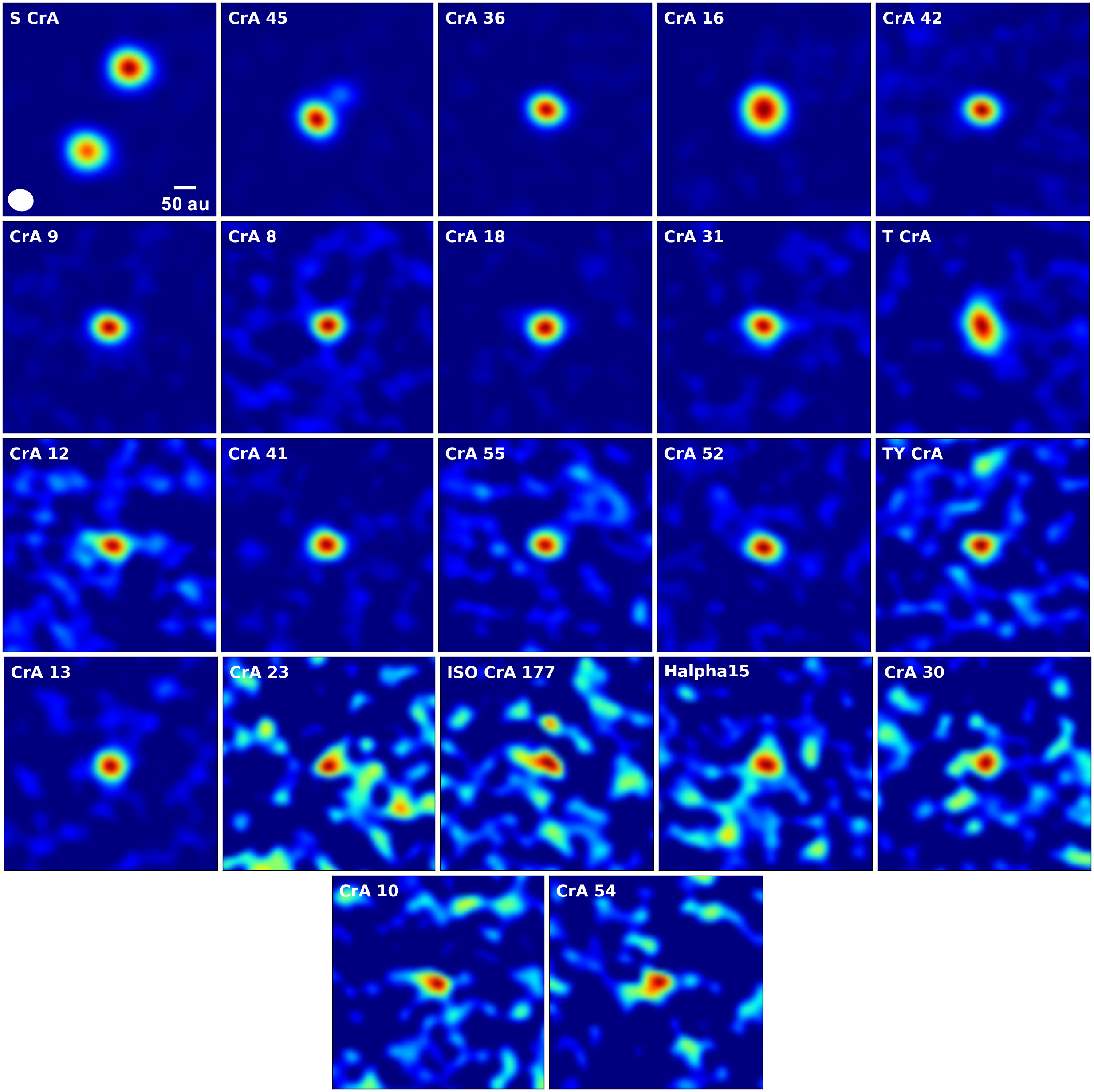}
\caption{ALMA Band 6 1.3 mm continuum images of the 24 detections in Corona Australis. The size of the images is $3\asec\times3\asec$. The size of the beam is indicated at the bottom-left corner of the first panel ($0\asec.31\times0\asec.33$). The north of each image is upwards. The presented images have not been primary beam corrected.}
\label{fig:all}
\end{figure*}

The continuum data imaging was performed with multi-frequency synthesis (MFS) imaging of the continuum data using the CASA-{\tt clean} task, and correcting for the primary beam.
By jointly imaging all three epochs of data, for each target source field, the achieved continuum root-mean-square (RMS) noise level is $\sim$0.15 mJy\,beam$^{-1}$, and the synthesized beam is \beam=0\farcs33$\times$0\farcs31 (P.A.=67$^{\circ}$), corresponding to a spatial resolution of $\sim50\,\rm au$ at $d=154\,\rm pc$. The imaged detections are presented in Fig. \ref{fig:all}.

It is important to note that because of an error when setting the observation coordinates, the decimal places of the target RAs have been trimmed: this results in an offset of the sources of up to 15$\asec$ east of the phase center: as a consequence, our images had to be primary beam corrected. The images in Fig.~\ref{fig:all} have therefore been re-centered using the best-fit positions in Tab.~\ref{tab:millimeter}.

\begin{table*}
  \caption{ALMA observations towards Class\,II objects in the CrA molecular cloud}\label{tab:observations}
  \begin{center}
    \begin{tabular}{p{5.5cm} |p{3.5cm} p{3.5cm} p{3.5cm} }
      \toprule
      Epoch & 1 & 2 & 3 \\
      Time (UTC; 2016)& (Aug.01) 03:32-04:54 & (Aug.01) 05:01-06:23 & (Aug.02) 03:18-04:40 \\
      \midrule
      Project baseline lengths (min-max) [m]& 14-1108 & 15-1075 & 15-1110 \\
      Absolute flux calibrator & Pallas & Pallas & Pallas \\
      Gain calibrator & J1937-3958 & J1924-2914 &  J1924-2914 \\
      Bootstrapped gain calibrator flux [Jy] & 0.26 & 3.9 & 4.1 \\
      Passband calibration & J1924-2914 & J1924-2914 & J1924-2914 \\
      Bootstrapped passband calibrator flux [Jy] & 4.1 & 3.9 & 4.1 \\
    \bottomrule
  \end{tabular}
  \end{center}
\end{table*}

\subsection{VLT/X-Shooter observations}\label{sec:data_xshooter}
The spectroscopic follow-up observations for the 13 targets with missing spectral type information were carried out in Pr.Id. 299.C-5048 (PI Manara) and Pr.Id. 0101.C-0893 (PI Cazzoletti) with the VLT/X-Shooter spectrograph \citep{Vernet11}. This instrument covers the wavelength range  from $\sim$300 nm to $\sim$2500 nm simultaneously, dividing the spectrum in three arms, the UVB ($\lambda\lambda\sim$ 300-550 nm), the VIS ($\lambda\lambda \sim$ 500-1050 nm), and the NIR ($\lambda\lambda\sim$ 1000-2500 nm). All targets were observed both with a narrow slit - 1.0\arcsec \ in the UVB, 0.9\arcsec \ in the VIS and NIR arms - leading to R$\sim$9000 and $\sim$10000, respectively, and a wide slit of 5.0\arcsec \ used to obtain an accurate flux calibration of the spectra. The log of the observations is reported in Table~\ref{tab:logXS}.  The spectra of all the observed targets are detected in the NIR arm, while only 5 targets are bright enough and not extincted too much to be detected also in the UVB arm. 

The reduction of the data was performed using the ESO X-Shooter pipeline 2.9.3 \citep{Modigliani10}. The pipeline performs the typical reduction steps, such as flat fielding, bias subtraction, order extraction and combination, rectification, wavelength calibration, flux calibration using standard stars observed in the same night. We extracted the 1D spectra from the 2D images produced by the pipeline using IRAF and then removed telluric absorption lines in the VIS and NIR arms using telluric standard stars observed close in time and airmass (see e.g., \citealt{Alcala14}). The S/N of the spectra at different wavelengths is reported in Table~\ref{tab:logXS}.

\begin{table*}
\center
\caption{1.3 mm continuum properties of the sources targeted in our sample.}\label{tab:millimeter}
\begin{threeparttable}
\begin{tabular}{lcccccccc}
\toprule
Name         &       $\Delta\alpha$ &     $\Delta\delta$ &$    F_{\rm 1.3 \, mm}   $&   RMS                 &$    a_{\rm maj}          $&$ a_{\rm min}        $&   PA                         &$   M_{\rm dust}                 $\\                    
             &       [$\asec$]      &     [$\asec$]      &     [mJy]                &   [mJy beam$^{-1}$]   &    [$\asec$]              &  [$\asec$]           &   [$^\circ$]                 &    [$M_\oplus$]                 \\  
\midrule  
CrA-1        &       ...            &     ...            &$    ...                 $&   0.10                &$   ...                   $&$ ...                $&$  ...                   $    &$   ...                 $\\
CrA-4        &       ...            &     ...            &$    ...                 $&   0.14                &$   ...                   $&$ ...                $&$  ...                   $    &$   ...                 $\\
CrA-6        &       ...            &     ...            &$    ...                 $&   0.08                &$   ...                   $&$ ...                $&$  ...                   $    &$   ...                 $\\
CrA-8        &       -0.14          &     0.39           &$    2.06   \pm   0.17   $&   0.16                &$   0.365  \pm    0.018   $&$ 0.31 \pm     0.014 $&$  81.90  \pm    13.3    $    &$   1.50\pm0.12               $\\             
CrA-9        &       -0.01          &     0.35           &$    5.07   \pm   0.16   $&   0.33                &$   0.389  \pm    0.008   $&$ 0.31 \pm     0.005 $&$  81.07  \pm     3.4    $    &$   3.70\pm0.12               $\\              
CrA-10       &       0.07           &     0.34           &$    0.65   \pm   0.11   $&   0.22                &$   0.481  \pm    0.119   $&$ 0.25 \pm     0.034 $&$  73.11  \pm     7.9    $    &$   0.48\pm0.15               $\\              
CrA-12       &       -0.05          &     0.42           &$    1.37   \pm   0.24   $&   0.10                &$   0.661  \pm    0.100   $&$ 0.32 \pm     0.030 $&$  90.67  \pm     4.8    $    &$   1.00\pm0.17               $\\              
CrA-13       &       0.06           &     0.27           &$    2.77   \pm   0.26   $&   0.21                &$   0.367  \pm    0.021   $&$ 0.35 \pm     0.019 $&$ 126.02  \pm    60.2    $    &$   2.02\pm0.19               $\\             
CrA-15       &       ...            &     ...            &$    ...                 $&   0.08                &$   ...                   $&$ ...                $&$  ...                   $    &$   ...                 $\\
CrA-16       &       -0.23          &     0.45           &$    20.34  \pm   0.53   $&   1.00                &$   0.478  \pm    0.008   $&$ 0.44 \pm     0.007 $&$  13.42  \pm    11.5    $    &$   14.84\pm0.39               $\\             
CrA-18       &       -0.20          &     0.52           &$    5.36   \pm   0.19   $&   0.35                &$   0.380  \pm    0.008   $&$ 0.32 \pm     0.006 $&$  93.93  \pm     5.3    $    &$   3.92\pm0.14               $\\              
CrA-21       &       ...            &     ...            &$    ...                 $&   0.08                &$   ...                   $&$ ...                $&$  ...                   $    &$   ...                 $\\
CrA-22       &       ...            &     ...            &$    ...                 $&   0.12                &$   ...                   $&$ ...                $&$  ...                   $    &$   ...                 $\\
CrA-23       &       -0.26          &     0.76           &$    0.35   \pm   0.15   $&   0.12                &$   0.401  \pm    0.120   $&$ 0.28 \pm     0.061 $&$ 108.44  \pm    24.8    $    &$   0.26\pm0.11               $\\             
CrA-26       &       ...            &     ...            &$    ...                 $&   0.11                &$   ...                   $&$ ...                $&$  ...                   $    &$   ...                 $\\
CrA-28       &       ...            &     ...            &$    ...                 $&   0.09                &$   ...                   $&$ ...                $&$  ...                   $    &$   ...                 $\\
CrA-29       &       ...            &     ...            &$    ...                 $&   0.11                &$   ...                   $&$ ...                $&$  ...                   $    &$   ...                 $\\
CrA-30       &       -0.36          &     0.61           &$    0.51   \pm   0.18   $&   0.09                &$   0.504  \pm    0.139   $&$ 0.27 \pm     0.046 $&$  96.59  \pm    10.6    $    &$   0.37\pm0.13               $\\             
CrA-31       &       -0.28          &     0.60           &$    2.82   \pm   0.19   $&   0.19                &$   0.416  \pm    0.019   $&$ 0.33 \pm     0.013 $&$  80.58  \pm     7.3    $    &$   2.05\pm0.14               $\\              
CrA-35       &       ...            &     ...            &$    ...                 $&   0.11                &$   ...                   $&$ ...                $&$  ...                   $    &$   ...                 $\\
CrA-36       &       -0.14          &     0.38           &$    12.9   \pm   0.21   $&   0.82                &$   0.384  \pm    0.004   $&$ 0.32 \pm     0.002 $&$  74.97  \pm     2.1    $    &$   9.41\pm0.15               $\\              
CrA-40       &       ...            &     ...            &$    ...                 $&   0.11                &$   ...                   $&$ ...                $&$  ...                   $    &$   ...                 $\\ 
CrA-41       &       -0.34          &     0.62           &$    2.80   \pm   0.19   $&   0.21                &$   0.384  \pm    0.017   $&$ 0.30 \pm     0.011 $&$  85.01  \pm     6.7    $    &$   2.04\pm0.14               $\\              
CrA-42       &       -1.00          &     0.34           &$    4.87   \pm   0.21   $&   0.34                &$   0.377  \pm    0.010   $&$ 0.31 \pm     0.007 $&$  81.88  \pm     4.9    $    &$   3.55\pm0.16               $\\              
CrA-45 E     &       -0.36          &      0.63          &$     29.82  \pm   0.35  $&   1.92                &$    0.400  \pm    0.003  $&$  0.34 \pm     0.002$&$   46.21  \pm     1.7   $    &$   21.76\pm0.26               $\\               
CrA-45 W     &       0.05           &      0.28          &$     6.36   \pm   0.34  $&   1.92                &$    0.393  \pm    0.013  $&$  0.33 \pm     0.009$&$   87.43  \pm     7.0   $    &$   4.64\pm0.25               $\\               
CrA-47       &       ...            &     ...            &$    ...                 $&   0.09                &$   ...                   $&$ ...                $&$  ...                   $    &$   ...                 $\\
CrA-48       &       ...            &     ...            &$    ...                 $&   0.12                &$   ...                   $&$ ...                $&$  ...                   $    &$   ...                 $\\
CrA-52       &       0.05           &     -0.27          &$    1.95   \pm   0.17   $&   0.16                &$   0.404  \pm    0.023   $&$ 0.29 \pm     0.012 $&$  70.00  \pm     5.4    $    &$   1.43\pm0.12               $\\              
CrA-53       &       ...            &     ...            &$    ...                 $&   0.10                &$   ...                   $&$ ...                $&$  ...                   $    &$   ...                 $\\
CrA-54       &       0.61           &     -0.14          &$    0.48   \pm   0.19   $&   0.09                &$   0.577  \pm    0.189   $&$ 0.26 \pm     0.044 $&$ 105.67  \pm     7.8    $    &$   0.35\pm0.14               $\\              
CrA-55       &       -0.26          &     0.24           &$    0.81   \pm   0.14   $&   0.10                &$   0.354  \pm    0.037   $&$ 0.29 \pm     0.025 $&$  96.62  \pm    19.4    $    &$   0.59\pm0.10               $\\             
CrA-56       &       ...            &     ...            &$    ...                 $&   0.10                &$   ...                   $&$ ...                $&$  ...                   $    &$   ...                 $\\ 
CrA-57       &       ...            &     ...            &$    ...                 $&   0.10                &$   ...                   $&$ ...                $&$  ...                   $    &$   ...                 $\\ 
S CrA S      &       -0.36          &      1.34          &$    129.53 \pm   2.09   $&   10.4                &$   0.451  \pm    0.005   $&$ 0.40 \pm     0.004 $&$  75.03  \pm     4.3    $    &$   94.51\pm1.52               $\\              
S CrA N      &       0.23           &      0.13          &$    140.30 \pm   2.00   $&   10.4                &$   0.439  \pm    0.004   $&$ 0.39 \pm     0.003 $&$  80.73  \pm     3.7    $    &$   102.36\pm1.46               $\\              
T CrA        &       -0.06          &     1.34           &$    4.99   \pm   0.37   $&   0.28                &$   0.568  \pm    0.033   $&$ 0.37 \pm     0.016 $&$  20.25  \pm     4.3    $    &$   3.64\pm0.27                $\\              
TY CrA      &       0.01           &     0.46           &$    0.91   \pm   0.18   $&   0.13                &$   0.362  \pm    0.044   $&$ 0.28 \pm     0.026 $&$  90.13  \pm    14.6    $    &$   0.66\pm0.13                $\\             
IRS10        &       ...            &     ...            &$    ...                 $&   0.08                &$   ...                   $&$ ...                $&$  ...                   $    &$   ...                  $\\   
Halpha15     &       -0.07          &     0.54           &$    0.69   \pm   0.22   $&   0.08                &$   0.529  \pm    0.128   $&$ 0.39 \pm     0.078 $&$ 142.60  \pm    27.2    $    &$   0.50\pm0.16                $\\             
ISO-CrA-177  &       -0.01          &     0.49           &$    0.52   \pm   0.17   $&   0.11                &$   0.535  \pm    0.146   $&$ 0.25 \pm     0.037 $&$  71.91  \pm     7.3    $    &$   0.38\pm0.13                $\\              
Haas17       &       ...            &     ...            &$    ...                 $&   0.11                &$   ...                   $&$ ...                $&$  ...                   $    &$   ...                  $\\
G09-CrA-9    &       ...            &     ...            &$    ...                 $&   0.09                &$   ...                   $&$ ...                $&$  ...                   $    &$   ...                  $\\
\bottomrule
\end{tabular}
\begin{tablenotes}
\small 
\item \textbf{Notes.} $\dagger$ Offset with respect to coordinates listed in Tab.~\ref{tab:stellarprop}.
\end{tablenotes}
\end{threeparttable}
\label{default}
\end{table*}

\section{Results and analysis}\label{sec:results}
\subsection{Stellar properties}\label{sec:stellarprop}

The spectral type for the targets were obtained from the literature (see Tab.~\ref{tab:stellarprop}) or from the VLT/X-Shooter spectra. The procedure used for the analysis of the X-Shooter spectra was as follows. First, we corrected the spectra for extinction using the values from the literature \citep{dunham, 2008ApJ...687.1145S,2011ApJ...736..137S} and the reddening law by \citet{cardelli} with $R_V$=3.1, as suggested by \citet{2008ApJ...687.1145S}. Then, we calculated the values of a number of spectral indices both at wavelengths in the VIS and the NIR arms, taken by those calibrated by \citet{riddick07}, \citet{jeffries07}, and \citet{2014ApJ...786...97H}, as in \citet{2017A&A...604A.127M}, and by \citet{testi00}, as in \citet{manara13a}. The spectral types derived from these indices are presented in Tab.~\ref{tab:indices} in Appendix~\ref{sec:appendix_spectra}. The spectral indices in the VIS arms are more reliable, and we select the spectral type from these indices when available. The observed spectra along with a template of the relative Spectral Types are presented in Fig.~\ref{fig:spectra}.

The spectral types are converted in effective temperatures ($T_{\rm eff}$) using the relation by \citet{2014ApJ...786...97H}. Stellar luminosity ($L_\star$) is obtained from the reddening-corrected $J$-band magnitudes and using the bolometric correction from \citet{2014ApJ...786...97H}, assuming for all the target the average distance of 154 pc calculated by \citet{2018ApJ...867..151D}. With this information, we have been able to plot our data on the HR diagram (Fig.~\ref{fig:hrdiagram}) and to estimate the stellar masses ($M_\star$) for all the targets using the evolutionary tracks by \citet{B15} for $M_\star<1.4 M_\odot$ and \citet{Siess00} for higher $M_\star$ and ages younger than 1 Myr . The stellar parameters for the targets are reported in Tab.~\ref{tab_stprop_app}. 

\subsection{mm continuum emission}
Among the 41 targets, 20 of them show a clear ($\geq4\sigma$) detection within a $1\asec$ radius from the nominal \textit{Spitzer} location from \citet{peterson}. In addition, CrA-42 and T CrA show a $\sim36\sigma$ and a $\sim22\sigma$ detection respectively at a slightly larger distance from their nominal \textit{Spitzer} positions ($1\asec.05$ for CrA-42 and $1\asec.34$ T CrA), and are also regarded as detections. S CrA is a known binary \citep{1993A&A...278...81R,1997ApJ...481..378G,2003A&A...397..675T}, and we detected millimeter emission associated with both binary components. CrA-45 is also identified as a binary. The total number of detections is therefore 24 out of the 43 targeted disks, so the detection rate is $\sim56\%$ .

None of the disks show clear substructures, no transition disk with cavities with radius $>25\,\rm au$ are found and all of them appear to be unresolved or marginally resolved: a Gaussian is therefore fitted to the detected sources (two Gaussians for the binaries) in the image plane using the \texttt{imfit} task in \texttt{CASA}. The task returns the total flux-density $F_{\rm1.3\, mm}$ of the source along with the statistical uncertainty, the FWHM along the semi-major ($a_{\rm maj}$) and semi-minor ($a_{\rm min}$) axis and the position angle (PA). The results of the fit are shown in Table \ref{tab:millimeter}\footnote{Note that the $F_{\rm1.3\, mm}$ uncertainty only includes the statistical uncertainty from the fit, and not the 10\% absolute flux calibration uncertainty.}The right ascension offset ($\Delta\alpha$) and the declination offset ($\Delta\delta$) with respect to the \textit{Spitzer} coordinates is also shown. The rms noise for the non-detections was calculated using the \texttt{imstat} task within a $1\asec$ radius centered at the \textit{Spitzer} coordinates; for the detection, it was calculated in an annular region centered on the source and with inner and outer radii equal to $2\asec$ and $4\asec$, respectively.

In order to constrain the average flux density of individually undetected sources, a stacking analysis was also performed. The images were centered at their \textit{Spitzer} coordinates (Table \ref{tab:stellarprop}) and then stacked. Even after the stacking, no detection was found and an average rms noise is $0.017\,\rm mJy\,beam^{-1}$, corresponding to a $3\sigma$ upper limit of $0.051\,\rm mJy$ is found assuming unresolved disks. However, it should be noted that the average offset between the disks and the \textit{Spitzer} positions, measured on the detections, are $<\Delta\alpha>=-0.13\asec$ and $<\Delta\delta>=0.47\asec$: it is therefore possible that the undetected sources did not overlap during the stacking, and that the upper limit is actually higher than that quoted.

\subsection{Dust masses}\label{sec:dustmasses}
Assuming that the observed sub-millimeter emission is optically thin and isothermal, the relation between the emitting dust mass ($M_{\rm dust}$) and the observed continuum flux at frequency $\nu$ ($F_\nu$) is as follows \citep{1983QJRAS..24..267H}:
\begin{equation}\label{eq:dustmass}
M_{\rm dust}=\frac{F_\nu\,d^2}{\kappa_\nu\,B_\nu(T_{\rm dust})}\approx2.19\times10^{-6}\,\Bigg(\frac{d}{160}\Bigg)^2F_{\rm 1.3\,mm}\,\rm [M_\odot],
\end{equation}
where $d$ is the distance of the object, $F_\nu$ is measured the flux-density, $B_\nu(T_{\rm dust})$ is the Planck function for a given dust temperature $T_{\rm dust}$ and $\kappa_\nu$ is the dust opacity at frequency $\nu$. To make the comparison with previous surveys easier, for the dust opacity $\kappa_\nu$ we follow the same approach of \citet{lupus}, assuming $\kappa_\nu=10\,\rm cm^2\,g^{-1}$ at 1000~GHz \citep{1990AJ.....99..924B} and scaling it to our frequency using $\beta=1$. The adopted value is therefore $\kappa_\nu=2.3\,\rm cm^2\,g^{-1}$ at $\nu=230\,\rm GHz$ (1.3 mm). In the right-hand side of Eq. \ref{eq:dustmass}, the distance $d$ is measured in pc and the flux density $F_{1.3\,\rm mm}$ is in mJy. For each object, the average distance of the cluster $d=154$ pc was used. For the dust temperature, we use a constant $T_{\rm dust}=20\,\rm K$ \citep{2005ApJ...631.1134A}, rather than the $T_{\rm dust}=25{\,\rm K}\times(L_{*}/L_\odot)^{0.25}$ relation based on two-dimensional continuum radiative transfer by \citet{andrews2013} and used in other works \citep[e.g.][]{2017AJ....154..255L}. We adopt this simplified approach with a single grain opacity and temperature for all the disks in the sample following the approach of \citet{lupus} and to facilitate the comparison with other star-forming regions (see Sec. \ref{sec:comparison}). Moreover, it should be noted that no dependence of the average dust temperature on the stellar parameters was found with the more detailed modelling by \citet{2017A&A...606A..88T} for the Lupus disks.

The dust masses of the disks in our sample are presented in Tab. \ref{tab:millimeter}, along with the relative uncertainty calculated from the flux uncertainty. Only 3 disks out of 24 detections have a dust content $\geq 10\,\rm M_\oplus$\footnote{5 sources in total have a dust content $\geq 10\,\rm M_\oplus$ if we also consider CrA-16 and CrA-36, which have dust masses only marginally below $10\,\rm M_\oplus$.} and large enough to form the cores of giant planets in the future. However, it is still possible that a similar amount of dust mass is hidden at the inner few region due to very high optical depth \citep[e.g.][]{ 2010ApJ...713.1143Z, 2017A&A...602A..19L,2018A&A...614A..98V}. Also note that very recent high angular resolution ALMA and VLA observations of disks are revealing that an important amount of dust is located in dense regions such as rings \citep[e.g.][]{2018ApJ...869L..41A}, which are optically thick at wavelengths around 1 mm \citep{2018ApJ...869L..46D}. When optically thin emission is detected, higher masses are estimated \citep{2016ApJ...821L..16C}.

The stacking of the non-detections gives an average $3\sigma$ upper limit corresponding to $0.036\,\rm M_\oplus$, about 3 Lunar masses.

\begin{figure}
\centering
\includegraphics[width=\columnwidth]{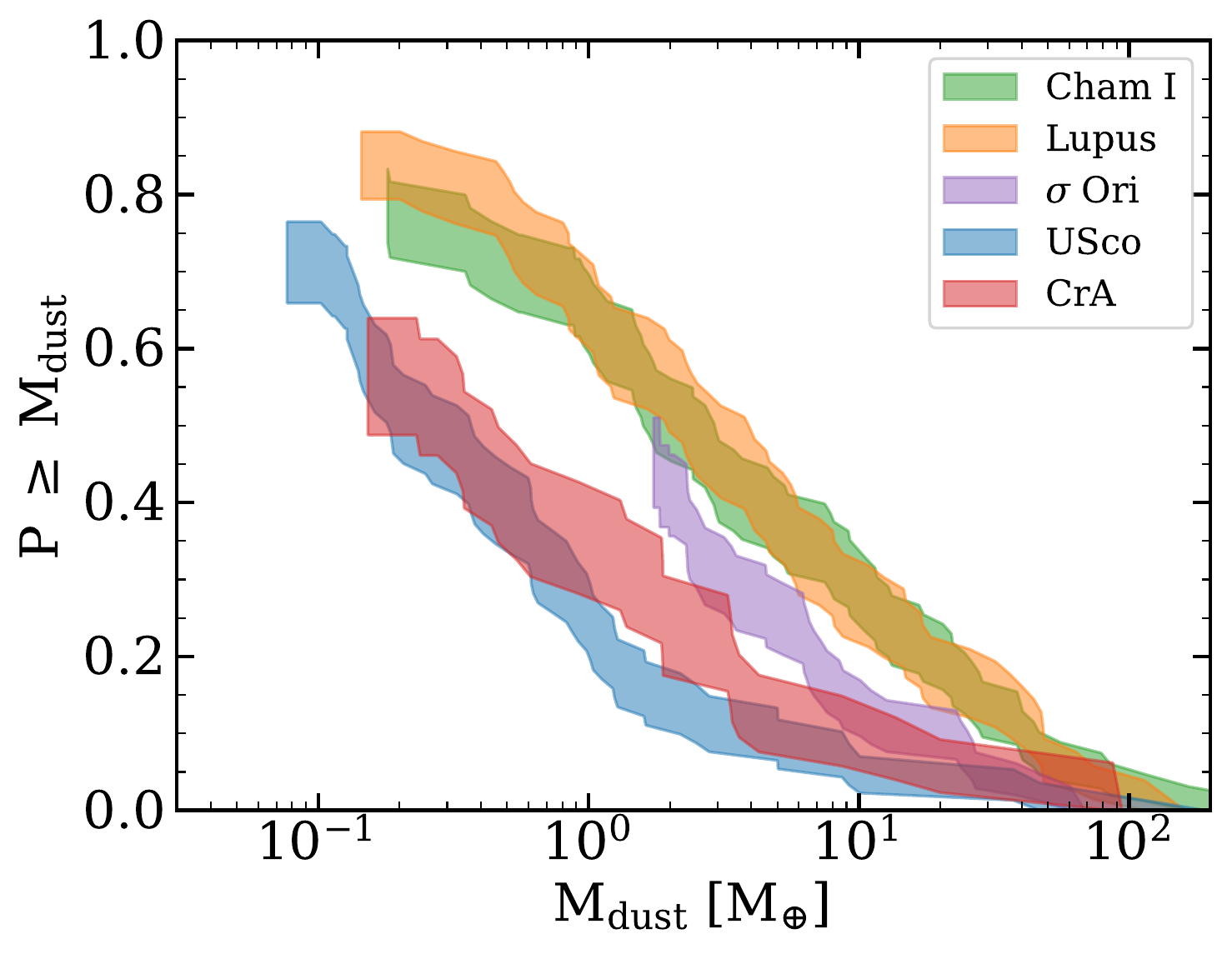}
\caption{
\footnotesize{
Comparison of the cumulative dust mass distributions of Lupus, CrA, Cham I, $\sigma$ Ori and Upper Sco, derived using a survival analysis accounting for the upper limits.
}\label{fig:cumulative}
}
\end{figure}

\subsection{Comparison with other regions}\label{sec:comparison}
The surveys of nearby star forming regions over the last years have shown growing evidence of a decrease in the mass of the disks with age, reflecting dust growth and disk dispersal. \Citet{lupus, 2017AJ....153..240A} found consistent results, calculating the highest average mass in the youngest regions (1-3 Myrs), and the lowest for the the oldest Upper Sco association (5-10 Myrs). The 2-3 Myrs old IC348 is the only exception, showing an average dust mass of only $4\pm1\,\rm M_\oplus$ (between the average $\sigma$ Orionis and that of Upper Sco) despite its young age. This can be explained by the  low-mass stellar population in the region \citep{2018MNRAS.478.3674R} (also see Tab.~\ref{tab:regions}).

\begin{table}
\center
\caption{Global properties of the star forming regions surveyed with ALMA in order of age.}\label{tab:regions}
\begin{threeparttable}
\begin{tabular}{lccc}
\toprule
Name         &       Distance &     Age        & Average dust mass\\
             &       [pc]     &     [Myr]      &     [$M_\oplus$]  \\
\midrule  
Taurus     &    129.5$^1$       &     1-3       &      $13\pm2$\\
Lupus     &    160$^{1,\star}$        &     1-3       &      $14\pm3$ \\  
CrA     &    154 $^1$        &     1-3       &      $6\pm3$ \\   
Chameleon I     &    192$^1$         &     2-3       &      $24\pm9$ \\ 
IC~348     &    321$^1$     &     2-3       &      $4\pm1$ \\   
$\sigma$ Ori     &    388$^1$         &     3-5       &      $7\pm1$ \\    
Upper Sco     &    144$^3$          &     5-10       &      $5\pm3$ \\

\bottomrule
\end{tabular}
\begin{tablenotes}
\small 
\item \textbf{References.} (1) \citet{2018ApJ...867..151D} (2) \citet{2008hsf2.book..295C} (3) \citet{1999AJ....117..354D}
\item \textbf{$\star$} The average distance of the 4 Lupus clouds was used.
\end{tablenotes}
\end{threeparttable}
\label{default}
\end{table}

The same analysis was done here for CrA. The dust masses are uniformly calculated following the approach used by \citet{lupus}, namely using Eq. \ref{eq:dustmass} with the continuum fluxes (or the $3\,\sigma$ upper limits) from our ALMA data or from the literature, assuming a uniform $T=20\,\rm K$, and inputting the frequency of the observation for each specific dataset. The distances assumed for each region are listed in Tab. \ref{tab:regions}. For the Upper Sco region, only the disks classified as "full", "evolved" and "transitional" from the \citet{2016ApJ...827..142B} sample are included, while the "debris" and Class III YSOs, which likely represent a separate evolutionary stage, are excluded. Finally, in order to facilitate the comparison with the other samples, in this analysis we only include the disks around stars with masses above the brown-dwarf limit ($M_\star\geq0.1\,\rm M_\odot$). The Kaplan-Meier estimator from the \texttt{lifelines}\footnote{10.5281/zenodo.1495175} and \texttt{ ASURV} \citep{1992ASPC...25..245L} packages were then used to estimate the cumulative mass distribution and to calculate the average dust mass and its uncertainty while properly accounting for the upper limits by using well-established techniques for left-censored data sets.

Fig.~\ref{fig:cumulative} presents the results accounting for the upper limits given by the non-detections. With an average dust mass of $6\pm3 \rm\, M_\oplus$, the distribution of the CrA disks appear closer to that of the old Upper Sco region rather than to those of the younger systems.
\subsection{$M_{\rm disk}-M_{\star}$ relation}\label{sec:mdiskmstar}
A clear correlation between the dust mass of disks and the mass of the central star has been identified across  all protoplanetary disk populations surveyed \citep{2016ApJ...831..125P,2017AJ....153..240A}. This finding highlights how the disk properties are affected by the central star, and is consistent with the correlation between frequency of giant planets and mass of the host star, both from the observational and theoretical points of view \citep{2011A&A...526A..63A, 2013A&A...549A.109B}. Moreover, the slope of this relation has been observed to steepen with time, with the young Taurus, Lupus and Chemeleon I regions ($\sim 1-3\,\rm Myr$) having slopes similar to each other and shallower than that found for the disks in the Upper Sco association ($5-10\,\rm Myr$).

\begin{figure}
\centering
\includegraphics[width=\columnwidth]{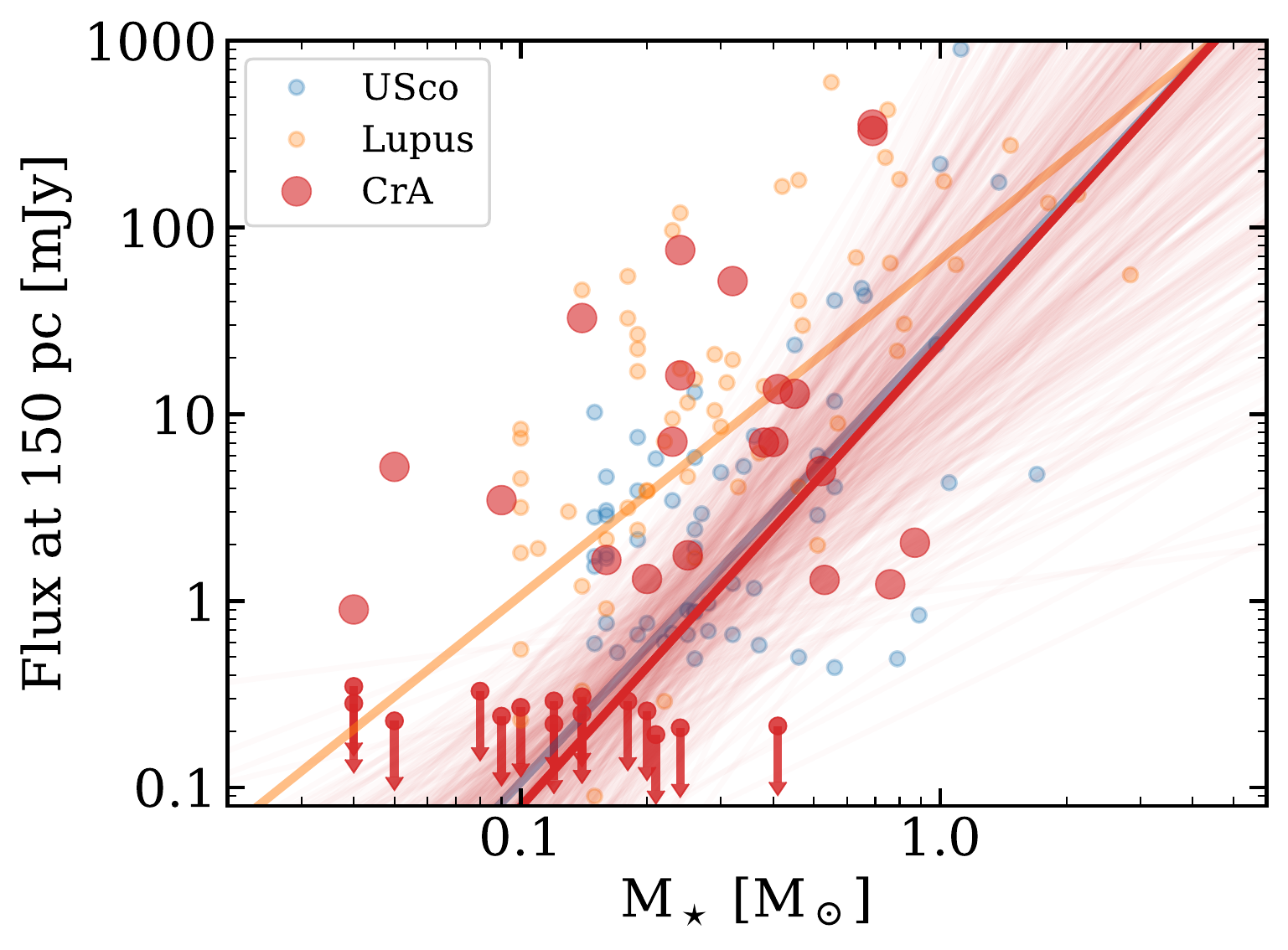}
\caption{
\footnotesize{
Correlation between dust disk flux scaled at 330 GHz \citep[assuming $\alpha=2.25$, as in][]{lupusb6} and at a distance of 150 pc with stellar mass for the objects in CrA. The slopes of Lupus and Upper Sco are also plotted for comparison. We show the results of the Bayesian fitting procedure by \citet{2007ApJ...665.1489K}. The solid line represents the best fit model, while the light lines show a subsample of models from the chains, giving an idea of the uncertainties.
}\label{fig:mdust_mstar}
}
\end{figure}

Studying the $M_{\rm dust}-M_\star$ relation for the disks in the CrA sample contributes to gain insight on the origin of the overall low mass of the disk population found in Sec. \ref{sec:comparison}. We derive the $M_{\rm dust}-M_\star$ relation using the same linear-regression Bayesian approach followed by \citet{2017AJ....153..240A} and presented by \citet{2007ApJ...665.1489K}\footnote{https://github.com/jmeyers314/linmix}. Unlike other linear regression methods, this approach is capable of simultaneously  accounting for the uncertainties in both the measurements of $M_{\rm dust}$ and $M_\star$, of the intrinsic scatter of the data and of the disk non detections, which result in upper-limits on the disk masses. Note that the SpT, and therefore the stellar mass, is missing for 5 of our targets: for these objects the stellar mass is randomly drawn from the stellar mass distribution of the entire sample. In particular, 4 of the objects with unknown SpT are also not detected with ALMA, while the other one (CrA-42) shows a clear detection of a disk at mm-wavelengths. For the 4 non detections, the stellar mass is therefore randomly drawn among the masses of the stars with non-detected disks, while the mass of CrA-42 is drawn from those showing a detection with ALMA. This uncertainty is also taken into account in the Bayesian approach we adopt by performing 100 different draws. In our fit, a standard uncertainty of 20\% of $M_\star$ on the stellar mass is assumed \citep{2017A&A...600A..20A,2017A&A...604A.127M}, while the uncertainties shown in Tab.~\ref{tab:millimeter} were used for the $M_{\rm dust}$ values. Finally, it should be noted that only 1 out of 89 sources in Lupus was a Herbig Ae/Be star, while Upper Sco did not include any Herbig. We therefore decided not to include {T CrA} and {TY CrA} in the fit , for which the $M_{\rm dust}-M_\star$ relation might not hold.

The best fit relation we find is then plotted in Fig. \ref{fig:mdust_mstar} in dark red, along with a subsample of all the models in the chains to show the uncertainty. As in the other surveys, we also find a correlation, where the best-fit model has a slope $\beta=2.32\pm0.77$ and intercept $\alpha=1.29\pm0.60$. This regression intercept is lower that that of other regions, as a consequence of the low disk masses found in the region. The uncertainties of the best-fit parameters reflect the large scatter in the data and the low number statistics.

In order to test that no strong bias was introduced by our procedure, we also run the fit described above without any random draw, finding consistent results.

\section{Discussion}\label{sec:discussion}

\subsection{Is CrA old?}\label{cra_old}
The observed low disk dust masses suggest that the CrA objects targeted in our survey may have an age comparable to that of the Upper Sco association, rather than to the young Lupus region. Unlike CrA, however, Upper Sco shows no presence of Class 0 or Class I sources, as expected for a $5-10$ Myr region \citep{dunham}. Moreover, most studies agree in assigning Corona Australis an age $<3\,\rm Myr$ \citep[e.g.][]{2009PASP..121..350M,2005A&A...429..543N,2008ApJ...687.1145S,2011ApJ...736..137S}.

On the other hand, most of these studies focused only on the \textit{Coronet} cluster, a small region extending  $\sim 1\,\rm pc$ around the {R CrA} YSO, and where most of the young embedded Class 0 and Class I sources are located (see Fig.~\ref{fig:map_classes}). The hypothesis that the large scale YSO population of the whole CrA cloud also includes a population of older objects therefore cannot be entirely ruled out. Some evidence of an additional older population has already been presented in previous studies. \citet{2000A&AS..146..323N} for example identify two classical T Tauri stars located outside the main cloud with an age of $\sim10\,\rm Myr$ using ROSAT data. In addition, \citet{peterson} perform a clustering analysis of the 116 YSOs in their sample, identifying a single core (corresponding to the \textit{Coronet}) and a more extended population of PMS stars showing an age gradient west of the \textit{Coronet}. They also observe that in the central core, the ratio {Class II/ Class I=1.8}, while the same ratio is {Class II/ Class I=2.3} when all the objects in the sample are considered, again hinting toward a younger population inside the \textit{Coronet}. Finally, \citet{2011ApJ...736..137S} point out that the relatively low disk fraction observed in the \textit{Coronet} \citep[$\sim 50\%$][based on near IR photometry]{2010A&A...515A..31L} is in strong contrast with the young age of the system: this inconsistency could be solved if an older population were also present. The large scatter in the $M_{\rm dust}-M_\star$ relation could also be a consequence of two stellar populations of different ages.

\begin{figure}
\centering
\includegraphics[width=\columnwidth]{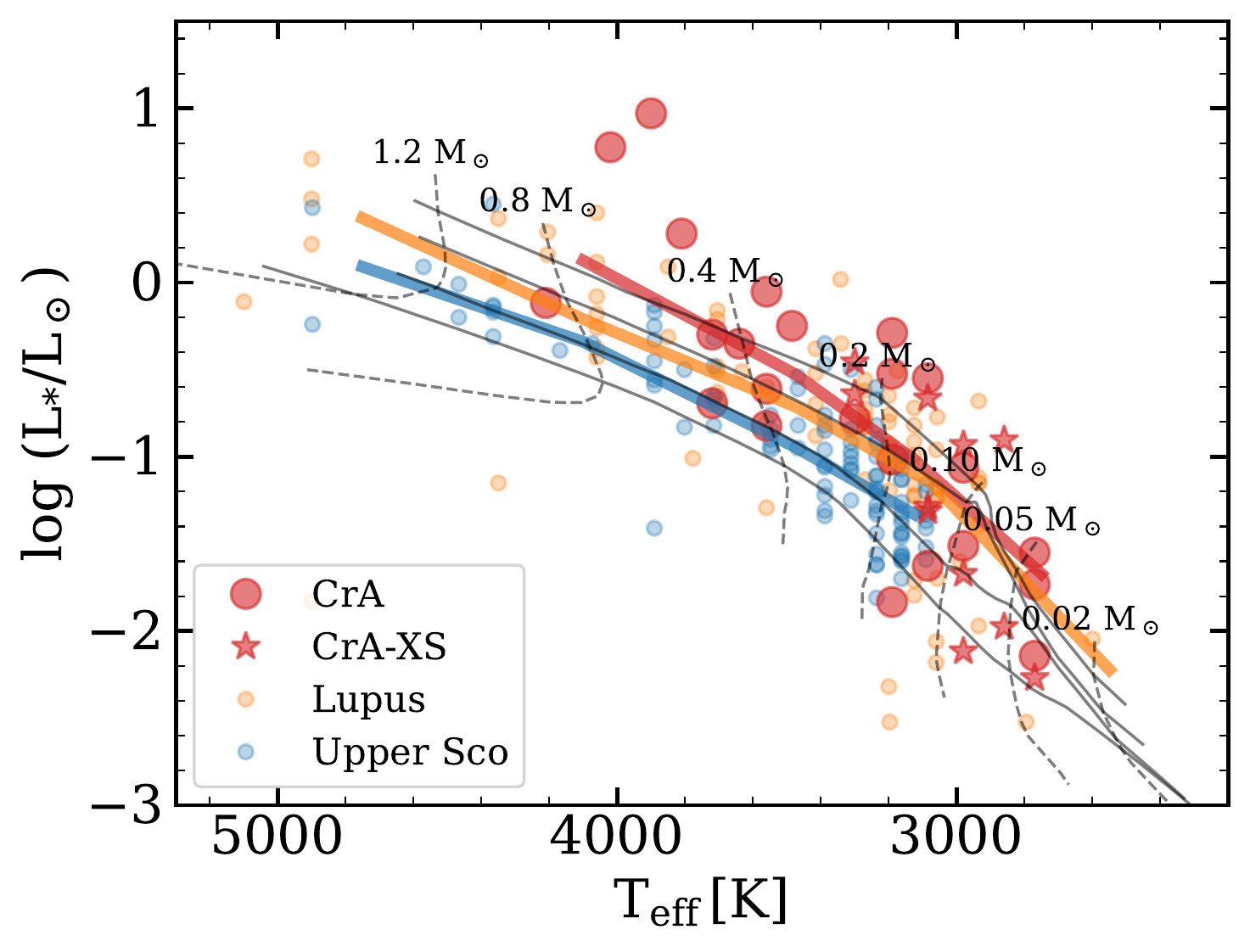}
\caption{
\footnotesize{
HR diagram showing the sources in our CrA sample (red), in the Lupus sample from \citet{lupus} (orange) and in the Upper Sco sample (blue) from \citet{2016ApJ...827..142B}. The evolutionary tracks for different stellar masses and the relative isochrones from \citet{B15} are also plotted for reference. The isochrones refer (from top to bottom) to the 1 Myr, 2Myr, 5 Myr and 10 Myr isochrone. The coloured solid lines show the approximate median value of the luminosity at each temperature. 
}\label{fig:hrdiagram}
}
\end{figure}

In order to further test if the Class II population in our sample indeed includes an older population, we have placed them on the HR diagram, by using the spectral types  listed in Tab. \ref{tab:stellarprop} and by deriving effective temperatures and bolometric corrections using the relationships in \citet{2014ApJ...786...97H} and tables in \citet{2015ApJ...808...23H}, respectively. The obtained diagram is presented in Fig. \ref{fig:hrdiagram}. For comparison, the Upper Sco and Lupus objects are also plotted. In contrast with what Fig. \ref{fig:cumulative} suggests, the HR diagram supports the scenario of a young CrA cluster with an age more consistent to that of Lupus than to Upper Sco.

In order to make this conclusion evident, the median values of the bolometric luminosities for each temperature are also shown (solid coloured lines in Fig. \ref{fig:hrdiagram}). The indicative age of the cluster is the isochrone closer to those median values: these lines also suggest that CrA is younger than Upper Sco. However, a more extended spectral classification for a larger number of objects in CrA would be needed to fully test this older-population scenario.

\subsection{Is CrA young?}
If the whole CrA is coeval with an age of $1-3\,\rm Myr$, some other mechanism has to be invoked to explain the low observed mm fluxes. For example, these fluxes could be due to low metallicity. However, \citet{2006A&A...446..971J} determined metallicities for three T Tauri stars in CrA, finding them to be only slightly sub-solar, and not low enough to explain our obesrvations.

External photo-evaporation is also known to play an important role in the disk mass evolution \citep{2016MNRAS.457.3593F,2018MNRAS.478.2700W}, and evidence of it occurring has been found in $\sigma\,\rm Ori$ \citep{2016ApJ...829...38M,2017AJ....153..240A}, where a clear correlation between disk mass and distance from the central Herbig O9V star has been observed and in the Orion Nebula Cluster \citep{2010ApJ...725..430M,2018ApJ...860...77E}. However, in CrA no correlation between the mass of the disks (or the disk detection rate) and the distance from the brightest star (R CrA) is found. Moreover, in $\sigma\,\rm Ori$ external photo-evaporation has been shown to affect disks up to 2 pc away from the Herbig star, where the geometrically diluted far-ultraviolet (FUV) flux reached a value of $\sim2000\,G_0$. The spectral type of R CrA is still uncertain, ranging from  F5 \citep[e.g.][]{2006A&A...459..837G} to B8 \citep[e.g.][]{2005ApJ...618..360H}. Even in the latter case, assuming a typical FUV luminosity for a B8 star of $L_{\rm FUV}\sim10\,\rm L_\odot$ \citep{2015A&A...582A.105A} and accounting for geometric dilution, we find that the FUV flux would drop to $\sim 1\, G_0$ in the first inner pc from R CrA, thus ruling-out external photo-evaporation as an explanation. Also,  this calculation neglects dust absorption, which is probably very effective in the Coronet cluster around R CrA.

\begin{figure}
\centering
\includegraphics[width=\columnwidth]{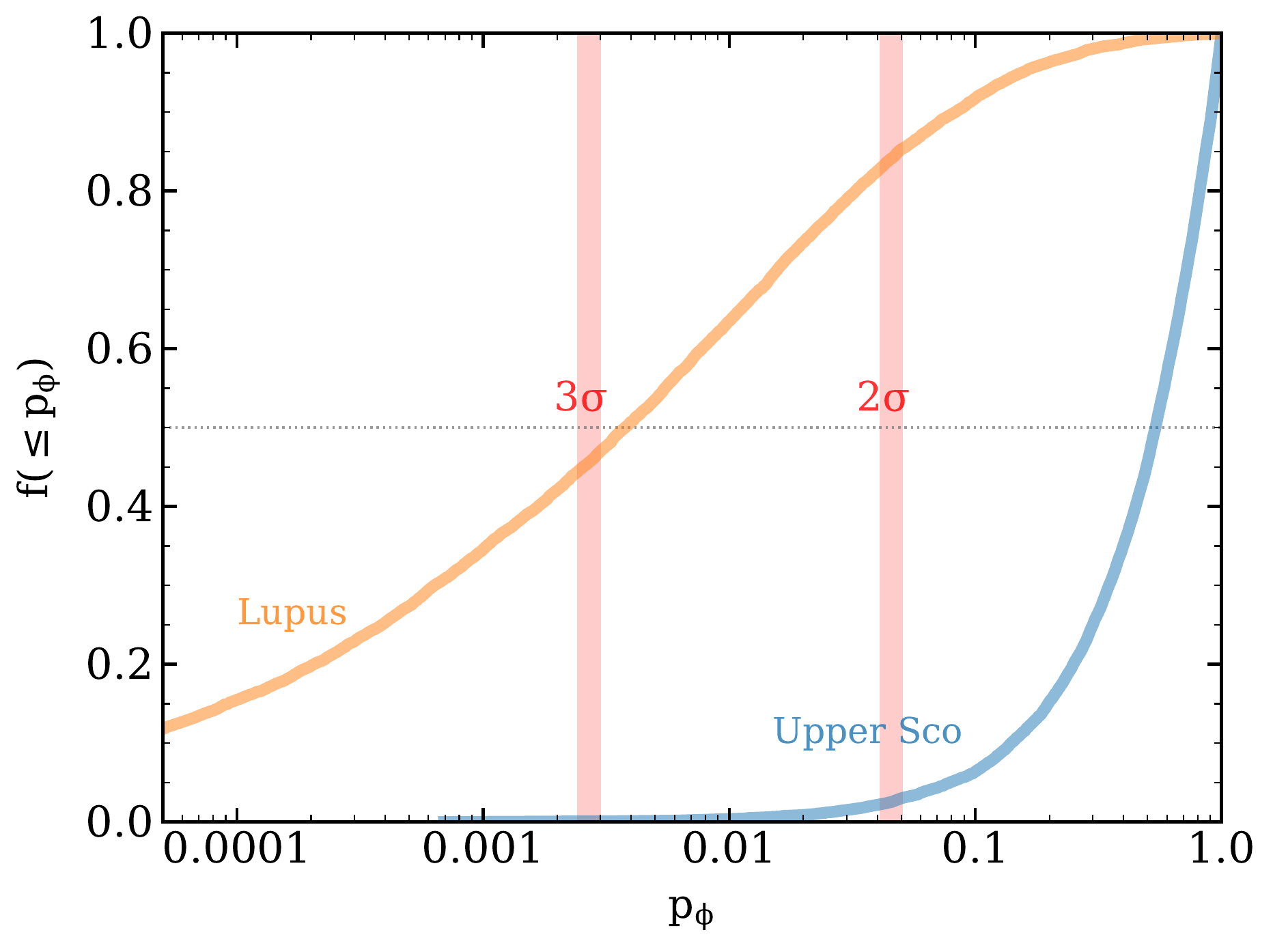}
\caption{
\footnotesize{
Comparison of the mass distributions of Lupus and USco to that of CrA, following the MC analysis proposed by \citet{andrews2013}. $p_\phi$ is the probability that the synthetic population drawn from the comparison sample (Lupus and Upper Sco) and the reference sample come from the same parent population.  $f(<p_\phi)$ is the cumulative distribution for $p_\phi$ resulting from the logrank two-sample test for censored datasets after $10^4$ MC iterations.
}\label{fig:mc_massdistr}
}
\end{figure}

Because of the $\rm M_{disk}-M_{\star}$ relation presented in Sec. \ref{sec:mdiskmstar}, it is also possible that a system dominated by low-mass stars shows a low-mass disk population, regardless of its age, as in the case of IC348 \citep{2018MNRAS.478.3674R}. It is therefore important, when comparing disk dust masses from different regions, to verify that they have the same stellar mass distribution. In order to do this, we employ a Monte Carlo (MC) approach similar to that used by \citet{andrews2013}. We first normalize the stellar populations by defining stellar mass bins and randomly drawing the same number of sources in each bin from the reference sample (CrA) and from a comparison sample (Lupus, Chamaeleon I or Upper Sco). We then perform a two-sample logrank test for censored datasets between the disk  dust masses of the two samples, to test the probability ($p_\phi$ value) that the two samples are randomly drawn from the same parent population. A low $p_\phi$ value indicates that the difference in disk masses cannot only be ascribed to different stellar populations and that some other factor, such as disk evolution and the age of the system, must play a role. This process is repeated $10^4$ times, and the results are used to create the cumulative distributions shown in Fig. \ref{fig:mc_massdistr}. When using Upper Sco as a comparison sample, we find a median $p_\phi$ value of 0.53, while the median $p_\phi$ for Lupus is only 0.004. The conclusion is that even when accounting for the $\rm M_{\rm dust}-M_\star$ relation, the disk dust mass distribution of CrA appears to be statistically different from that of Lupus, while it is significantly more similar to that from that of Upper Sco. Therefore, the comparably low masses of the protoplanetary disks in CrA cannot be explained in terms of the low stellar masses.

Another way a disk can lose part of its mass is via tidal interaction with other stars \citep[e.g.][]{1993MNRAS.261..190C,2005A&A...437..967P}. This mechanism is, however, only effective in much denser environments than CrA \citep[e.g.][]{2018MNRAS.475.2314W}. In principle, it is possible to imagine that at very early stages most of the stars were located in a dense region (e.g. the \textit{Coronet}) where they interacted violently before being ejected. However, the very low velocity dispersion of the stars in the cluster makes this scenario very unlikely \citep{2000A&AS..146..323N}. Tidal interaction can be effective in removing dust mass from a disk even in later stages when the disk is in a binary system \citep[e.g.][] {1994ApJ...421..651A}, as proposed to explain the low mm flux of some objects in Taurus by \citet{2018ApJ...869...17L}. A higher than usual binary fraction could therefore explain the low disk masses observed in CrA. However, \citet{1997ApJ...481..378G} show that the binary fraction of CrA is indistinguishable from those of Lupus and Chameleon I.

Finally it is possible that the low mass distribution observed today is a consequence of a population of disks that has formed with a low mass from the very beginning. For example, the disk formation efficiency in a cloud with mass $M_0$ depends on the sound speed $c_{\rm s}$ and on the solid body rotation rate $\Omega_0$, where we have defined the disk formation efficiency as the fraction of $M_0$ that is in the disk at the end of the collapse stage, or as the ratio between $M_{\rm disk}/M_\star$ at that time \citep{1981Icar...48..353C, 1984ApJ...286..529T}. In particular, clouds with higher $c_{\rm s}$ and  $\Omega_0$ (i.e. warmer or more turbulent) will form more massive disks \citep[also see Appendix A in][]{2009A&A...495..881V}. Therefore, a cold parent cloud or one with low intrinsic angular momentum $\Omega_0$, will form disks with a lower mass, and with a lower $M_{\rm disk}/M_\star$ as observed in CrA. Consistently, observations of dense cloud cores in the CrA cloud show line-widths lower than in other regions \citep{2002A&A...385..909T}.  Moreover, because of the smaller circularization radius, the formed disks would alse be  smaller \citep[e.g.][]{2006ApJ...645L..69D} and potentially mostly optically thick, thus hiding an even larger fraction of the mass . 
Alternatively, small and optically thick disks could result from magnetic braking of the disks by means of the magnetic field threading the disk and the surrounding molecular cloud at the formation stage \citep[e.g.][]{2008ApJ...681.1356M,2014ApJ...786...97H,2013ApJ...767L..11K}. The same scenario was proposed by \citet{2019A&A...621A..76M} to explain the low occurrence of large ($>60\,\rm AU$) Class 0 disks in the CALYPSO sample.

Such scenarios, although not testable with the present dataset, are consistent with the low disk mass distribution and with the low intercept of the $\rm M_{disk}-M_{\star}$ in CrA and are not in contradiction with  the young age of the stellar popultion. If the parent cloud initial conditions are indeed responsible for the low masses observed, this would be an additional critical aspect to be considered when studying planet formation and evolution. Since the conditions at the epoch of disk formation can be different in each star-forming region, proper modelling is required to assert to which extent they can affect the initial disk mass distribution, the subsequent disk evolution, planet formation and planetary populations. 

Observationally, this could be tested by observing the mass of disks around Class 0 and Class I objects in CrA: if the disks are born with a low-mass, the disk mass distribution even at these younger stages should be significantly lower than in other regions.

\section{Conclusion}\label{sec:conclusion}
We presented the first ALMA survey of 43 Class II protoplanetary disks in the Corona Australis nearby ($d=160\,\rm pc$) star forming region, in order to measure their dust content and understand how it scales with the stellar properties. The ultimate goal was  to test if the relations between disk properties, age of the stellar population found in other surveys also hold for this region.
\begin{enumerate}
\item The average mm fluxes from the disks in CrA is low. This in turn converts into a low disk mass distribution. Even though our observations are able to constrain dust masses down to $\sim0.2\,\rm M_\oplus$, the detection rate is only $56\%$. Moreover, we find that only 3 disks in our sample have a dust mass $\geq 10\,\rm M_\oplus$ and thus sufficient mass to form giant planet cores.
\item We obtained VLT/X-Shooter spectra for 8 objects with previously unknown spectral type, and derived their stellar physical properties.
\item Despite the apparent young age of the CrA stellar population, we find that the dust mass distribution of the disks in CrA is much lower than that of the Lupus young star forming region which shares a similar age, while it appears to be consistent with that in the 5-10 Myr old Upper Sco association. The correlation between disk dust mass $\rm M_{\rm dust}$ and stellar mass $\rm M_\star$ previously identified in all other surveyed star forming regions is confirmed. However, because of the low mass of the disks in our sample we find a much lower intercept. The large scatter of the data points  does not allow the slope of the relation to be well constrained for CrA.
\item Since most of the age estimates of the CrA regions are based on the population of the compact \textit{Coronet} cluster, a possible explanation for the low disk masses might be in principle that CrA also hosts an old population of disks, consistently with previous observations. The position of the objects of our sample on the HR diagram, however, seems to support the idea of a mostly coeval, young population.
\item Low disk masses in a young star forming region can be explained by external photo-evaporation (as in the case of $\sigma$ Ori) or by a low stellar mass population (as in IC348). With our analysis, we can rule out both these scenarios for CrA. Tidal interaction between different members of CrA, stripping material from the disks, as well as close binaries  can also be ruled out.
\item We suggest that initial conditions may play a crucial role in setting the initial disk mass distribution and its subsequent evolution. Small disks with low mass can originate from a cloud with very low turbulence or sound speed, or can alternatively result from disk magnetic braking. It is therefore important to better study the impact of initial conditions on  the disk properties, especially if planet formation occurs even before 1 Myr age, as the recent results from  \citet{2018ApJS..238...19T}  and \citet{2018A&A...618L...3M} suggest.
\end{enumerate}
Future surveys including younger Class 0 and I objects in CrA and other star forming regions will help testing wether or not initial conditions play a critical role in shaping the physical properties of circumstellar disks.

\begin{acknowledgements}
We thank S. van Terwisga, S. Andrews, G. Lodato and A. Hacar for very useful discussion, and Dr. Mark Gurwell for compiling the SMA Calibrator List (http://sma1.sma.hawaii.edu/callist/callist.html). We also acknowledge the DDT Committee and the Director of the La Silla and Paranal Observatory for granting DDT time. This work was partly supported by the Italian Ministero dell’Istruzione, Università e Ricerca through the grant Progetti Premiali 2012 – iALMA (CUP C52I13000140001), by the Deutsche Forschungs-gemeinschaft (DFG, German Research Foundation) - Ref no. FOR 2634/1 TE 1024/1-1, and by the DFG cluster of excellence Origin and Structure of the Universe (www.universe-cluster.de). This project has received funding from the European Union’s Horizon 2020 research and innovation programme under the Marie Skłodowska-Curie grant agreement No 823823. H.B.L. is supported by the Ministry of Science and
Technology (MoST) of Taiwan (Grant Nos. 108-2112-M-001-002-MY3 and 108-2923-M-001-006-MY3). J.M.A. acknowledges financial support from the project PRIN-INAF 2016 The Cradle
of Life—GENESIS-SKA (General Conditions in Early Planetary Systems for the rise 
of life with SKA). C.F.M. and S.F. acknowledge an ESO Fellowship. M.T. has been supported by the DISCSIM project, grant agreement 341137 funded by the European Research Council under ERC-2013-ADG and by the UK Science and Technology research Council (STFC). Y.H. is supported by the Jet Propulsion Laboratory, California Institute of Technology, under a contract with the National Aeronautics and Space Administration. C.C.G and R.G.M acknowledge financial support from DGAPA UNAM. This paper makes use of the following ALMA data: ADS/JAO.ALMA\#2015.1.01058.S. 
ALMA is a partnership of ESO (representing its member states), NSF (USA) and NINS (Japan), together with NRC (Canada) and NSC and ASIAA (Taiwan), in cooperation with the Republic of Chile. 
The Joint ALMA Observatory is operated by ESO, AUI/NRAO and NAOJ. All the figures
were generated with the \texttt{python}-based package \texttt{matplotlib} \citep{hunter2007}.
\end{acknowledgements}

\bibliographystyle{aa}

\appendix

\clearpage

\section{Additional stellar properties}

Tab. \ref{tab_stprop_app} shows a compilation of the most relevant stellar parameters used in our analysis. The $J$ magnitude is taken from the 2MASS survey \citep{2003yCat.2246....0C}. The extinctions are either derived from our VLT/X-Shooter spectra or from the references in Column 4. Note that the extinctions from \citet{dunham} were not derived from the stellar spectra but from extinction maps and might therefore systematically overestimate the real extinction towards the star by 1-2 mag \citep[see Sec.~ 4.1 in][]{peterson}. In order to make sure that the extinction values from \citet{dunham} were accurate, we compared them to those derived from the spectra by \citet{2011ApJ...736..137S} for 8 targets common to the two samples. We found that the extinctions derived with the two methods are consistent within the uncertainties. The effective temperatures and bolometric luminosities have finally been derived as explained in Sec. \ref{cra_old} and then used to determine the stellar masses as explained in Sec.~\ref{sec:stellarprop}.

\begin{table*}
\center
\caption{Compilation of the most relevant stellar properties used in our analysis. Only the stars with known Spectral Type were included.}\label{tab_stprop_app}
\begin{threeparttable}
\begin{tabular}{lcccccc}
\toprule
Source         &   J       &  $A_{\rm v}$  &   Ref.        &    $T_{\rm eff}$ &      $\log L_\star/L_\odot$ &     $M_\star$           \\                                      
             &   [mag]   &  [mag]        &               &    [K]          &                              &     [$M_\odot$]         \\            
\midrule          
CrA-1       &    10.99  &   0.0          &   1           &    2860         &       -0.90                  &     0.10$^{\rm s}$      \\                                               
CrA-4       &    13.98  &   3.3          &   2           &    2770         &       -1.73                  &     0.04$^{\rm b}$      \\                                    
CrA-6       &    10.77  &   2.2          &   2           &    3190         &       -0.52                  &     0.21$^{\rm b}$      \\                                    
CrA-8       &    12.92  &   1.2          &   2           &    2770         &       -1.55                  &     0.05$^{\rm b}$      \\                                    
CrA-9       &    10.38  &   2.1          &   2           &    3720         &       -0.29                  &     0.45$^{\rm b}$      \\                                    
CrA-10      &    14.19  &   2.7          &   3           &    3190         &       -1.83                  &     0.16$^{\rm b}$      \\                                   
CrA-12      &    12.95  &   1.4          &   2           &    2980         &       -1.51                  &     0.09$^{\rm b}$      \\                                    
CrA-13      &    12.83  &   8.1          &   3           &    3560         &       -0.61                  &     0.38$^{\rm b}$      \\                                   
CrA-15      &    14.85  &   14.0         &   3           &    3300         &       -0.78                  &     0.24$^{\rm b}$      \\                                   
CrA-16      &    14.45  &   17.0         &   3           &    3485         &       -0.24                  &     0.32$^{\rm b}$      \\                                   
CrA-18      &    13.90  &   14.0         &   3           &    3640         &       -0.35                  &     0.41$^{\rm b}$      \\                                   
CrA-21      &    14.91  &   13.5         &   3           &    3560         &       -0.82                  &     0.41$^{\rm b}$      \\                                   
CrA-22      &    12.33  &   1.1          &   1           &    3085         &       -1.28                  &     0.14$^{\rm b}$      \\                                         
CrA-23      &    14.08  &   0.08         &   3           &    2770         &       -2.14                  &     0.04$^{\rm b}$      \\                                   
CrA-26      &    15.58  &   4.2          &   1           &    2770         &       -2.26                  &     0.04$^{\rm b}$      \\                                         
CrA-28      &    13.41  &   1.9          &   3           &    3085         &       -1.62                  &     0.12$^{\rm b}$      \\                                   
CrA-30      &    9.31   &   3.3          &   2           &    3810         &        0.28                  &     0.53$^{\rm s}$      \\                                     
CrA-31      &    10.59  &   2.0          &   1           &    3300         &       -0.45                  &     0.23$^{\rm b}$      \\                                        
CrA-35      &    12.03  &   2.1          &   2           &    2980         &       -1.06                  &     0.12$^{\rm b}$      \\                                   
CrA-36      &    14.57  &   12.1         &   1           &    2980         &       -0.93                  &     0.14$^{\rm b}$      \\                                        
CrA-40      &    11.61  &   4.0          &   1           &    3085         &       -0.66                  &     0.18$^{\rm b}$      \\                                        
CrA-41      &    10.46  &   4.7          &   3           &    3560         &       -0.05                  &     0.40$^{\rm b}$      \\                                  
CrA-45      &    11.91  &   5.0          &   1           &    3300         &       -0.63                  &     0.24$^{\rm b}$      \\                                        
CrA-47      &    13.67  &   0.0          &   1           &    2860         &       -1.97                  &     0.05$^{\rm b}$      \\                                       
CrA-48      &    14.06  &   0.0          &   1           &    2980         &       -2.11                  &     0.08$^{\rm b}$      \\                                       
CrA-52      &    10.82  &   0.2          &   3           &    3720         &       -0.69                  &     0.52$^{\rm b}$      \\                                 
CrA-53      &    13.38  &   1.5          &   1           &    2980         &       -1.67                  &     0.09$^{\rm b}$      \\                                       
CrA-54      &    7.60   &   1.4          &   3           &    4020         &        0.77                  &     0.76$^{\rm s}$      \\                                    
CrA-55      &    9.78   &   1.0          &   3           &    4210         &       -0.11                  &     0.87$^{\rm b}$      \\                                  
CrA-56      &    12	    &   2.2          &   3           &    3190         &       -1.01                  &     0.20$^{\rm b}$      \\                                  
CrA-57      &    12.31  &   0.8          &   1           &    3085         &       -1.31                  &     0.14$^{\rm b}$      \\                                        
SCrA N      &    8.49*  &   7.9          &   2           &    3900         &        0.97                  &     0.69$^{\rm s}$      \\                                      
SCrA S      &    8.49*  &   7.9          &   2           &    3900         &        0.97                  &     0.69$^{\rm s}$      \\                                      
TCrA        &    8.93   &   7.9          &   2           &    7200         &        1.46                  &     2.25$^{\rm s}$      \\                                 
TYCrA       &    7.49   &   7.9          &   2           &    10500        &        2.47                  &     4.10$^{\rm s}$      \\                             
Halpha15    &    11.82  &   0.8          &   4           &    3190         &       -0.28                  &     0.25$^{\rm s}$      \\                                    
ISO-CrA-177 &    12.44  &   0.5          &   5           &    3085         &       -0.54                  &     0.20$^{\rm s}$      \\      

\bottomrule
\end{tabular}
\begin{tablenotes}
\small 
\item \textbf{$\bf A_{\rm v}$ references.} (1) This work (2)  \citet{dunham} (3) \citet{2011ApJ...736..137S} (4) \citet{1998ASPC..154.1755P}  (5) \citet{2005A&A...444..175L}
\item \textbf{Evolutionary Tracks.} (s) \citet{Siess00} (b) \citet{B15}
\end{tablenotes}
\end{threeparttable}
\label{default}
\end{table*}

\section{VLT/X-Shooter Spectra}\label{sec:appendix_spectra}

In this section we present the VLT/X-Shooter spectra obtained {Fig.~\ref{fig:spectra}). The Spectral Types derived from the different spectral indices calibrated are presented in Tab.~\ref{tab:indices}. In particular, SpT VIS was obtained from the average values from \citet{riddick07}, as in \citet{manara13a,2017A&A...604A.127M}; SpT TiO was obtained with the index by \citet{jeffries07}; SpT NIR was obtained with the indices by \citet{testi00}, as in \citet{manara13a}; the uncertainties represent the spread between the different indices in the VIS and NIR arms. The adopted spectral types, reported in the last column of Tab.~\ref{tab:indices}, are taken from the indices calculated in the VIS arm of the spectrum.
In addition, the log of the our X-Shooter observations is presented in Tab.~\ref{tab:logXS} along with the Signal to Noise ratio achieved at different wavelengths. 

\begin{table*}[h!]
\center
\caption{Spectral types derived from different spectral indices}\label{tab:indices}
\begin{threeparttable}
\begin{tabular}{lccc|c}
\toprule

Source    &  SpT VIS          &  SpT TiO      &    SpT NIR                  &       SpT       \\                
\midrule                  
CrA-1     &  M6.05$\pm$1.3    &    M5.46      &    M7.70$\pm$1.3            &        M6       \\            
CrA-22    &  M3.74$\pm$2.1    &    M4.48      &    M5.54$\pm$2.1            &        M4.5     \\              
CrA-26    &  M6.68$\pm$1.7    &    M0.64      &    L1   $\pm$1.7            &        M7$^\star$       \\           
CrA-31    &  M3.66$\pm$3.0    &    M3.61      &    M7.96$\pm$3.0            &        M3.5     \\             
CrA-36    &  M4.86$\pm$2.3    &    M2.84      &    L1   $\pm$2.3            &        M5$^\star$       \\            
CrA-40    &  M3.07$\pm$1.6    &    ...        &    M6.42$\pm$1.6            &        M4.5$^\star$     \\             
CrA-42    &  M4.44$\pm$3.5    &    ...        &    L2   $\pm$3.5            &        ...      \\             
CrA-45    &  M3.56$\pm$1.3    &    M2.22      &    M5.37$\pm$1.3            &        M3.5     \\             
CrA-47    &  M5.74$\pm$2.0    &    M5.87      &    L0.92$\pm$2.0            &        M6       \\            
CrA-48    &  M3.17$\pm$2.1    &    ...        &    M5.22$\pm$2.1            &        M5$^\star$       \\           
CrA-53    &  M5.02$\pm$1.1    &    M5.13      &    M7.90$\pm$1.1            &        M5       \\            
CrA-57    &  M4.05$\pm$1.8    &    M4.57      &    M5.89$\pm$1.8            &        M4.5     \\              
IRS10     &  ...              &    ...        &    L1.87$\pm$5.4            &        ...      \\            
                  
\bottomrule
\end{tabular}
\begin{tablenotes}
\item $\bf \star$ Uncertain estimate of SpT due to the low S/N of the spectra.
\end{tablenotes}
\end{threeparttable}
\end{table*}

\begin{table*}[h!]
\center
\caption{Night log and basic information on the spectra. In Column 1 is the name of the source, in Column 2 the date and time of the observations, in Column 3-5 the exposure times, in Column 6-8 the slit widths, in Column 9-11 the S/N measured at the indicated wavelengths, in Column 12-13 we show whether or not the $H_\alpha$ and Li lines have been detected.}\label{tab:logXS}
\begin{tabular}{l|c|ccc|ccc|ccc|cc}
\toprule

Source    & Date of observation [UT]  &  \multicolumn{3}{c}{Exp. Time [$N_{\rm exp}\times$ s]}      &    \multicolumn{3}{c}{Slit width [$\asec$]}         &      \multicolumn{3}{c}{S/N @ $\lambda$ [nm]}   & $H_\alpha$ & Li        \\                
          &                           &  UVB  & VIS & NIR                                          & UVB  & VIS & NIR                                     & 400              & 700    & 1000                &            &           \\ 
\midrule
\multicolumn{13}{c}{Pr.Id. 299.C-5048 (PI Manara)} \\
\midrule
CrA-31    &  2017-09-01T03:30:30.048  & 4x215   & 4x135 & 4x3x75                                   & 1.0  & 0.9 & 0.9                                     &  8               & 20     &   21               &   Y         & Y         \\   
CrA-36    &  2017-09-17T02:22:50.221  & 4x600   & 4x690 & 4x3x250                                  & 1.0  & 0.9 & 0.9                                     &  0               &  0     &   23               &   Y         & N         \\   
CrA-42    &  2017-09-09T02:14:39.978  & 4x630   & 4x700 & 4x3x250                                  & 1.0  & 0.9 & 0.9                                     &  0               &  0     &    1               &   N         & N         \\   
CrA-45    &  2017-09-06T00:37:53.044  & 4x440   & 4x340 & 4x3x150                                  & 1.0  & 0.9 & 0.9                                     &  0               & 24     & 1110               &   Y         & Y         \\   
\midrule
\multicolumn{13}{c}{Pr.Id. 0101.C-0893 (PI Cazzoletti)} \\
\midrule          
CrA-1     &  2018-05-28T04:25:20.708  & 4x90    & 4x150 & 4x150                                    & 1.0  & 0.9 & 0.9                                     & 24               & 14     & 35                 &   Y         &    Y        \\               
CrA-22    &  2018-06-14T06:36:52.359  & 4x190   & 4x250 & 4x250                                    & 1.0  & 0.9 & 0.9                                     &  3               & 20     & 58                 &   Y         &    ...        \\   
CrA-26    &  2018-06-12T05:54:41.945  & 4x190   & 4x250 & 4x250                                    & 1.0  & 0.9 & 0.9                                     &  0               &  2     & 13                 &   Y         &    N        \\   
CrA-40    &  2018-05-26T08:08:42.457  & 4x190   & 4x250 & 4x250                                    & 1.0  & 0.9 & 0.9                                     &  1               & 56     & 69                 &   Y         &    Y        \\   
CrA-47    &  2018-06-15T04:01:23.919  & 4x190   & 4x250 & 4x250                                    & 1.0  & 0.9 & 0.9                                     &  1               & 13     &309                 &   Y         &    Y        \\   
CrA-48    &  2018-05-26T05:05:48.019  & 4x190   & 4x250 & 4x250                                    & 1.0  & 0.9 & 0.9                                     & 11               & 21     & 20                 &   Y         &    N        \\   
CrA-53    &  2018-05-26T06:41:48.012  & 4x190   & 4x250 & 4x250                                    & 1.0  & 0.9 & 0.9                                     &  1               & 14     & 52                 &   Y         &    Y        \\   
CrA-57    &  2018-05-27T05:18:45.651  & 4x90    & 4x150 & 4x150                                    & 1.0  & 0.9 & 0.9                                     &  3               & 17     & 83                 &   Y         &    Y        \\   
IRS10     &  2018-06-11T06:08:20.787  & 4x190   & 4x250 & 4x250                                    & 1.0  & 0.9 & 0.9                                     &  0               &  0     &  2                 &   N         &    N        \\   
                  
\bottomrule
\end{tabular}
\end{table*}

\begin{figure*}[h!]
\begin{subfigure}{0.86\textwidth}
  \centering
  \includegraphics[width=\linewidth]{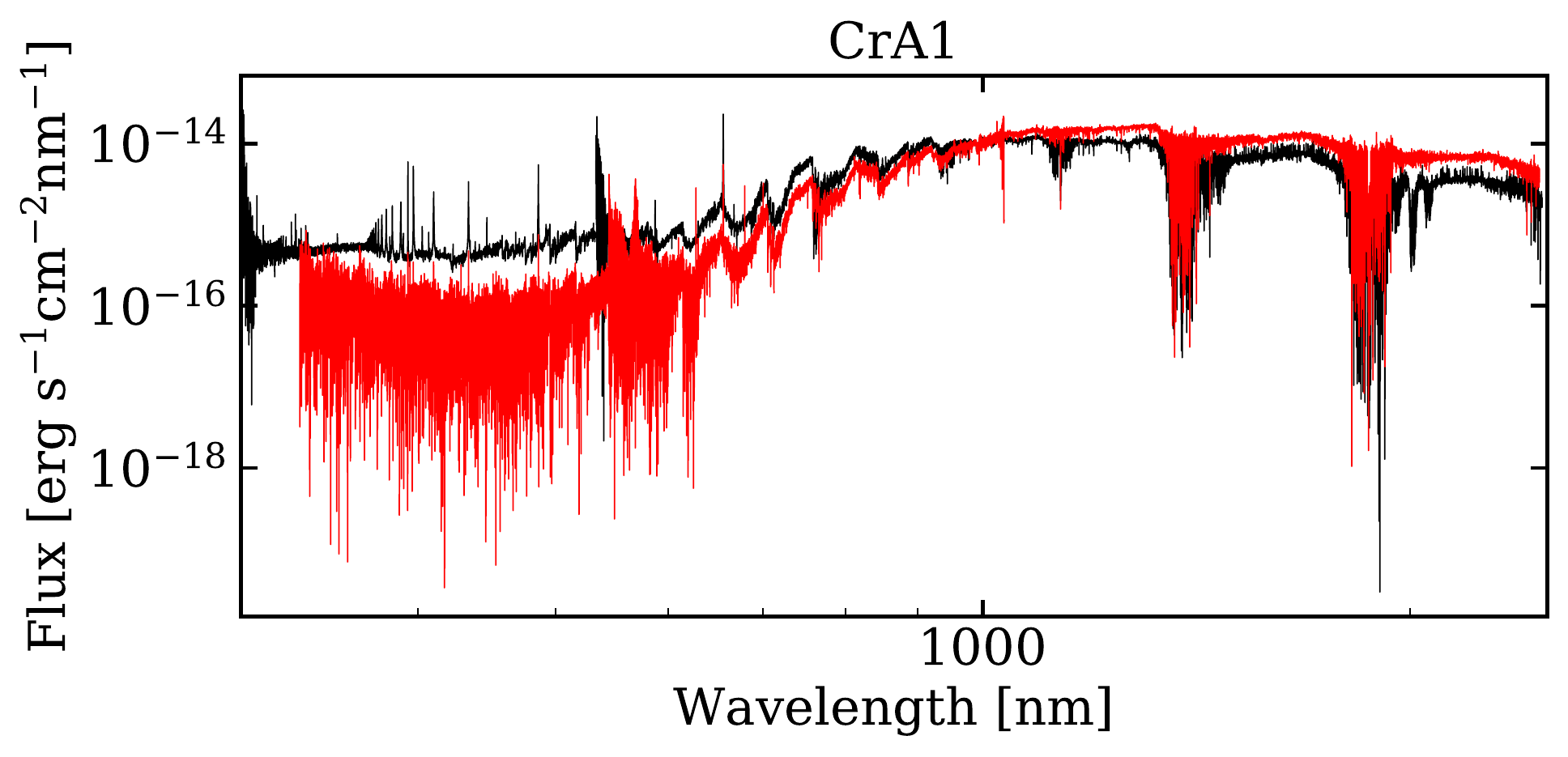}
\phantomcaption
\end{subfigure}%

\begin{subfigure}{.86\linewidth}
  \centering
  \includegraphics[width=\linewidth]{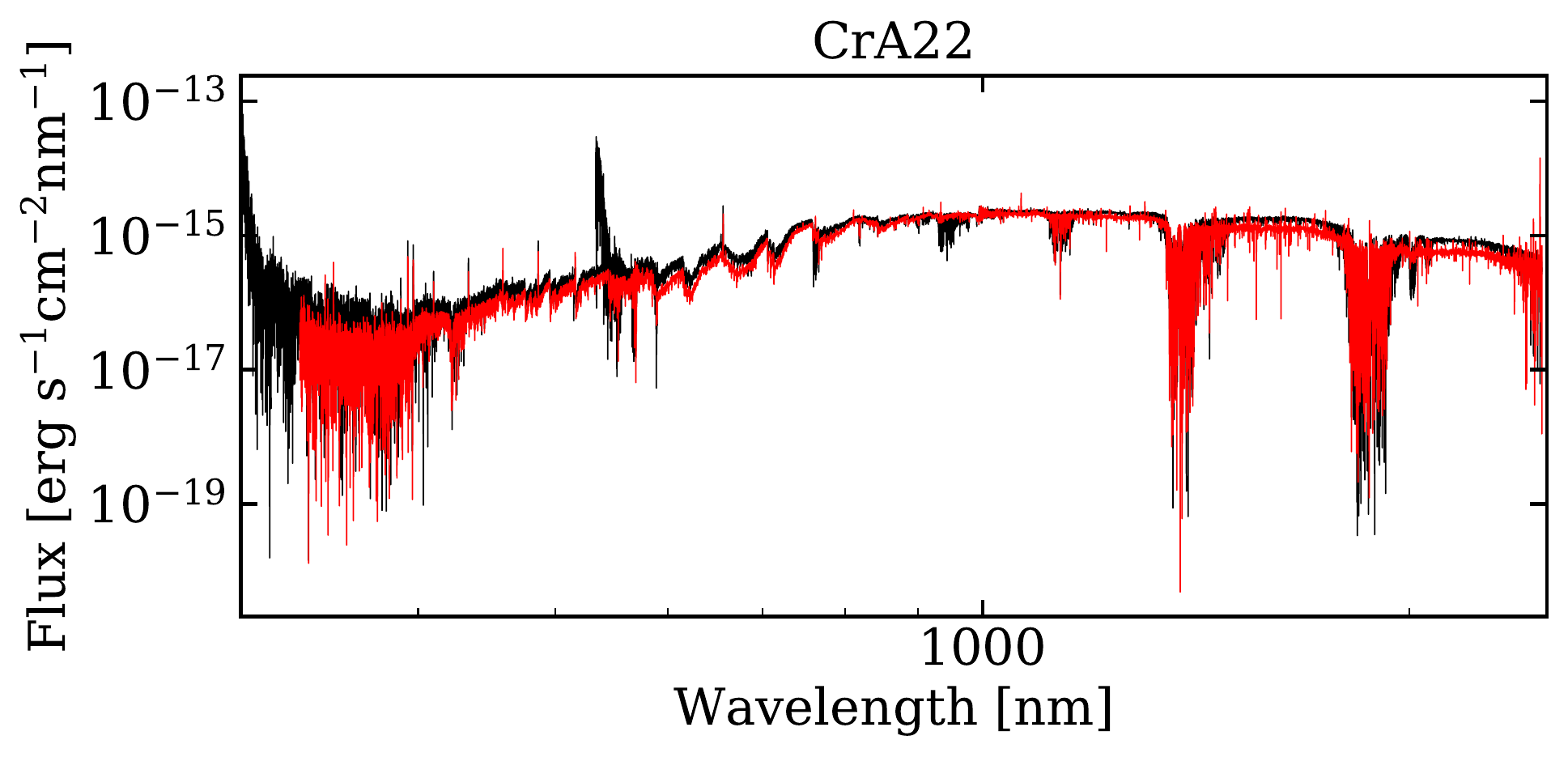}
\phantomcaption
\end{subfigure}

\begin{subfigure}{.86\textwidth}
  \centering
  \includegraphics[width=\linewidth]{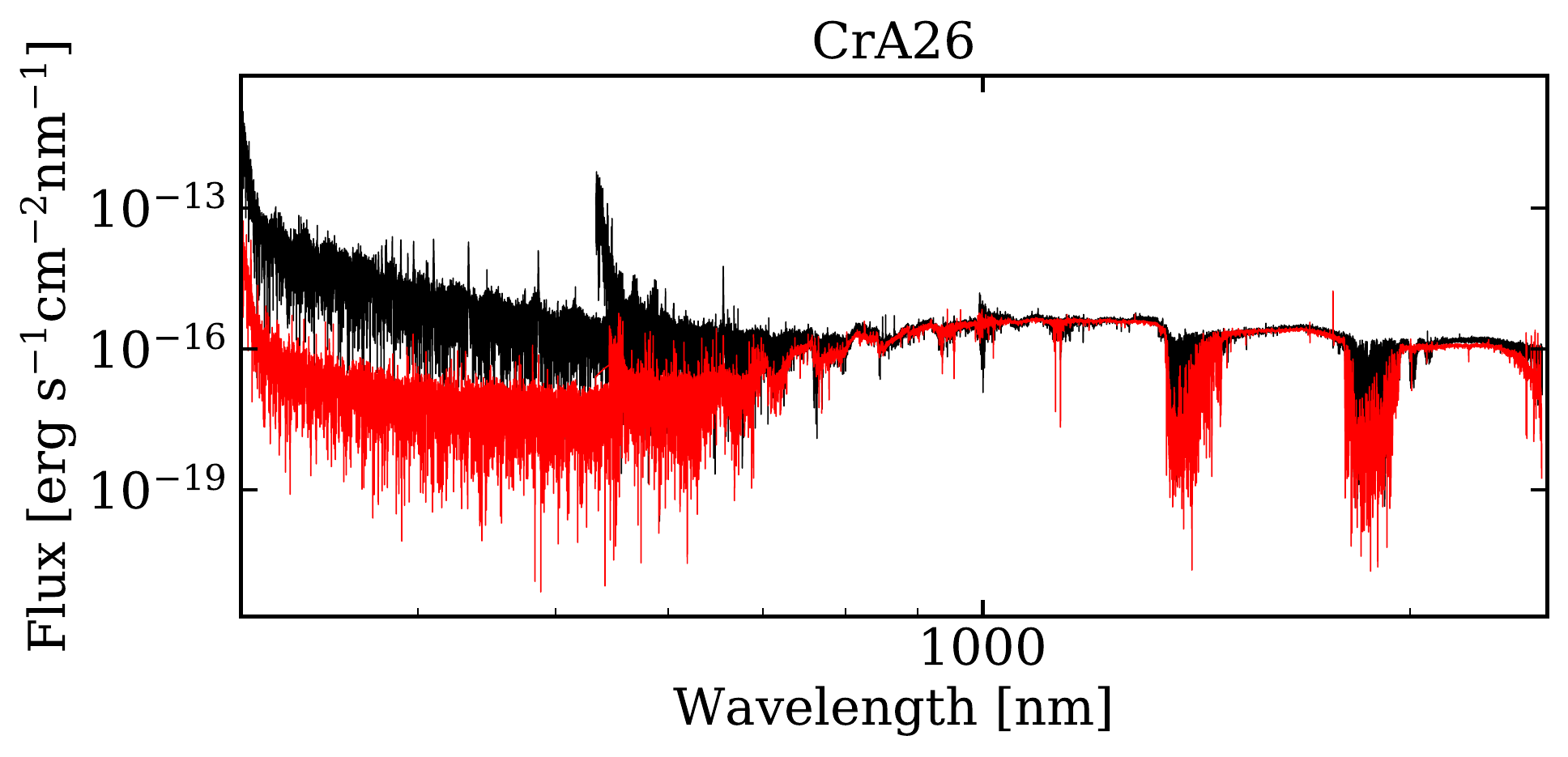}
\phantomcaption
\end{subfigure}
\caption{Spectra observed in our X-Shooter programs (black) along with a template with the same Spectral Type (red). The name of the sources is in the title of each subfigure. The absolute flux of each template was normalized to the flux of the observation at $\lambda=1\mu m$. }
\end{figure*}
\clearpage

\begin{figure*}\ContinuedFloat
\begin{subfigure}{0.86\textwidth}
  \centering
  \includegraphics[width=\linewidth]{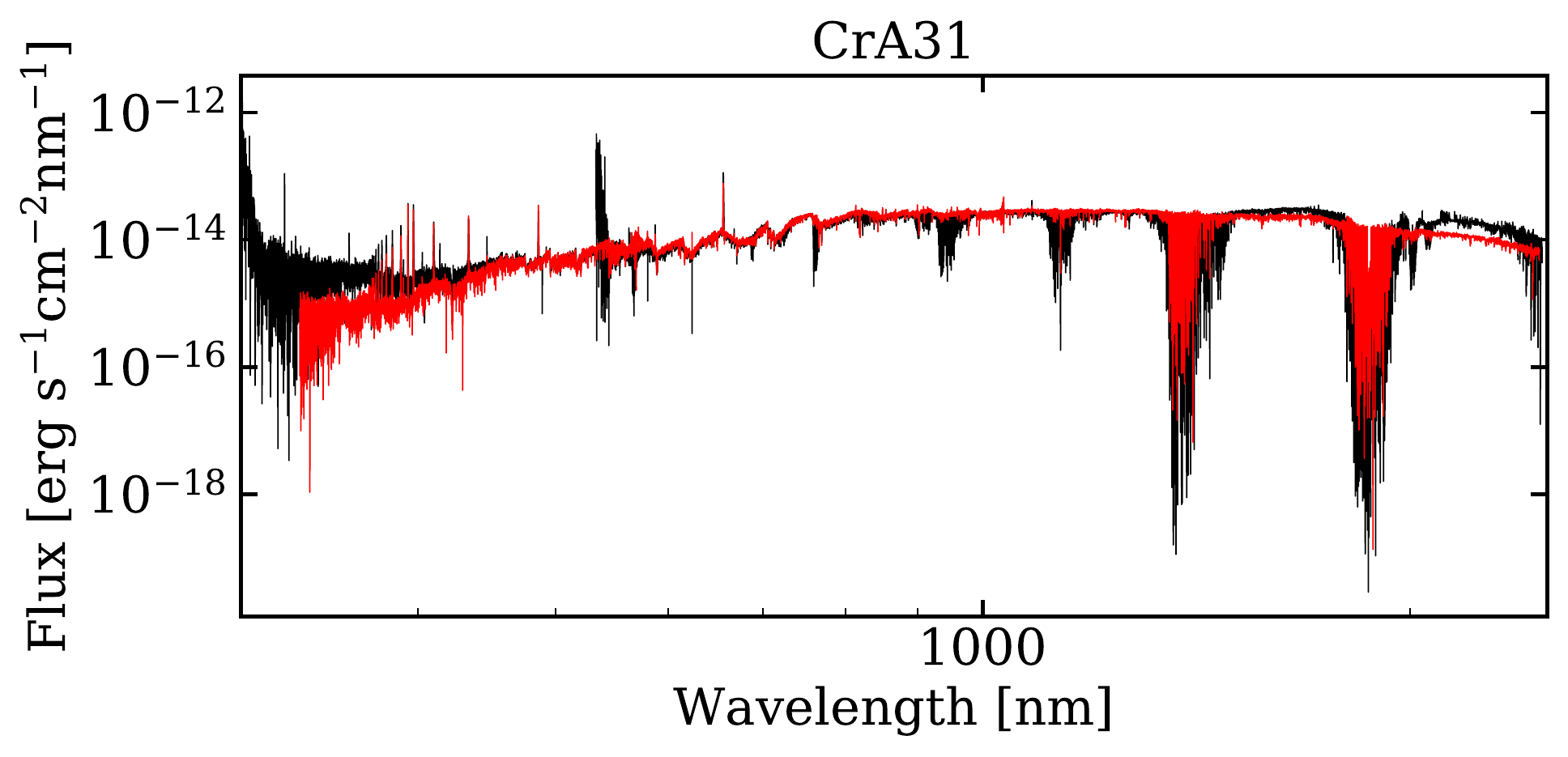}
\phantomcaption
\end{subfigure}%

\begin{subfigure}{.86\linewidth}
  \centering
  \includegraphics[width=\linewidth]{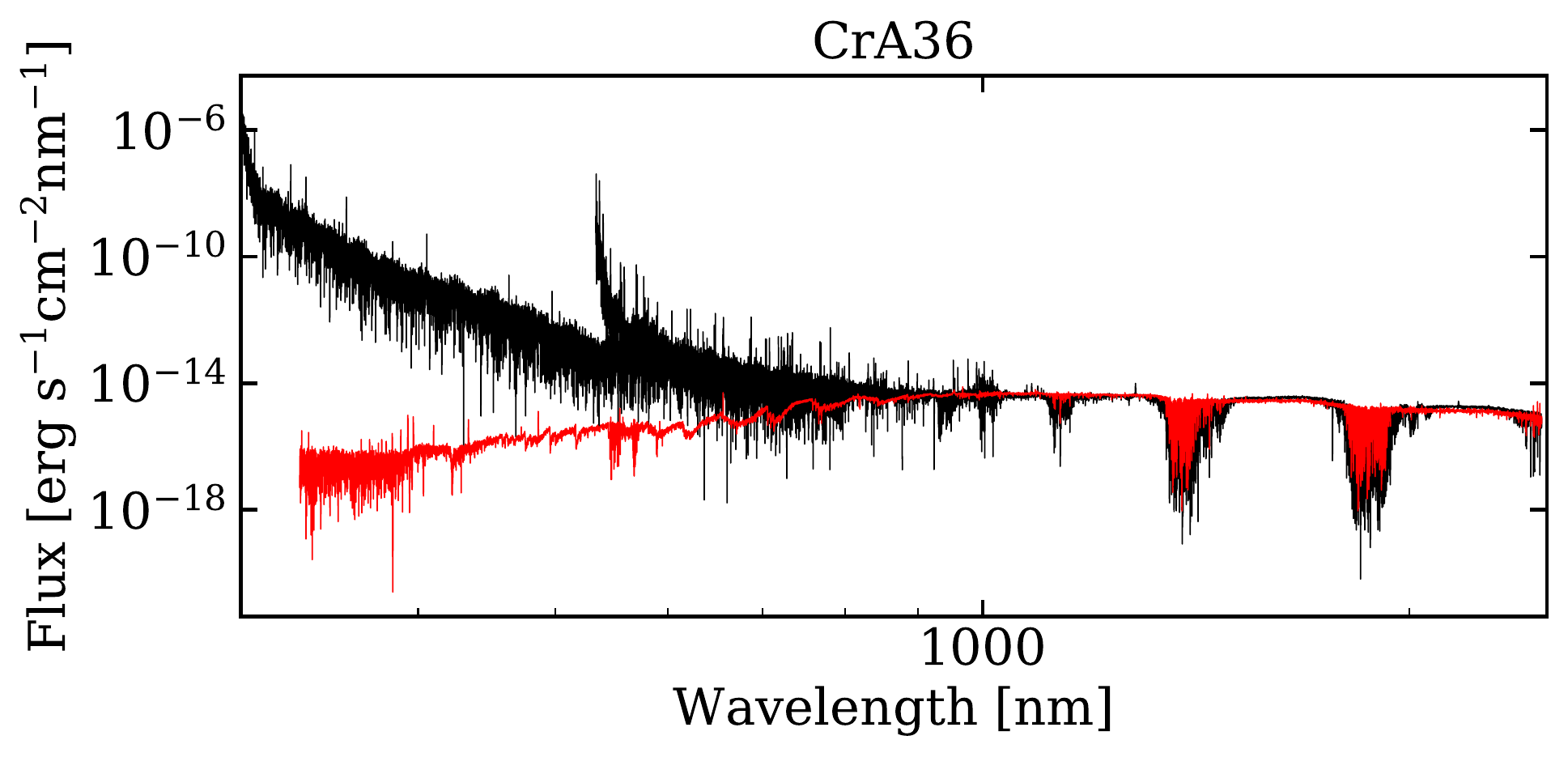}
\phantomcaption
\end{subfigure}

\begin{subfigure}{.86\textwidth}
  \centering
  \includegraphics[width=\linewidth]{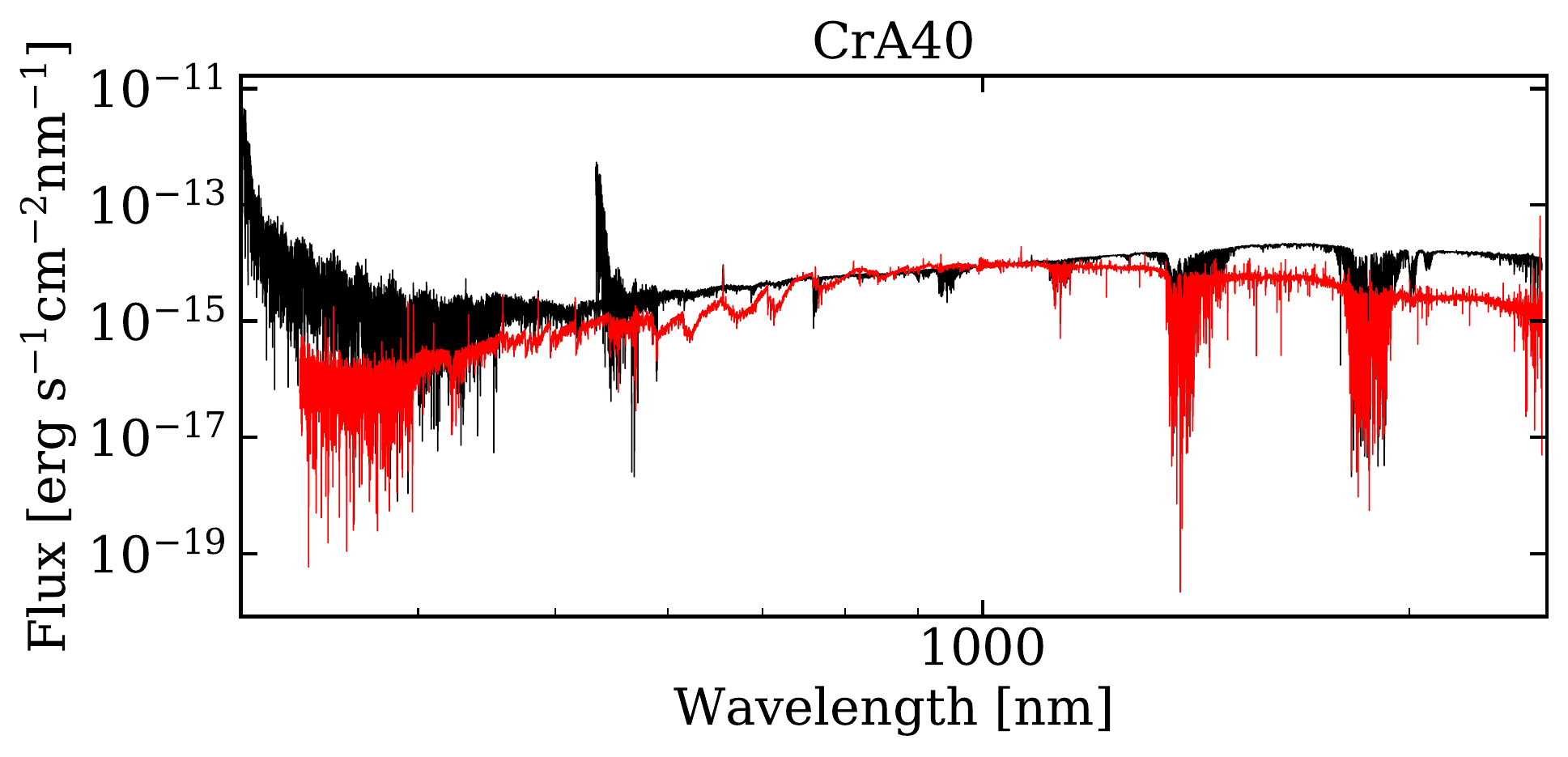}
\phantomcaption
\end{subfigure}
\caption{Continued}
\end{figure*}

\clearpage

\begin{figure*}\ContinuedFloat
\begin{subfigure}{0.86\textwidth}
  \centering
  \includegraphics[width=\linewidth]{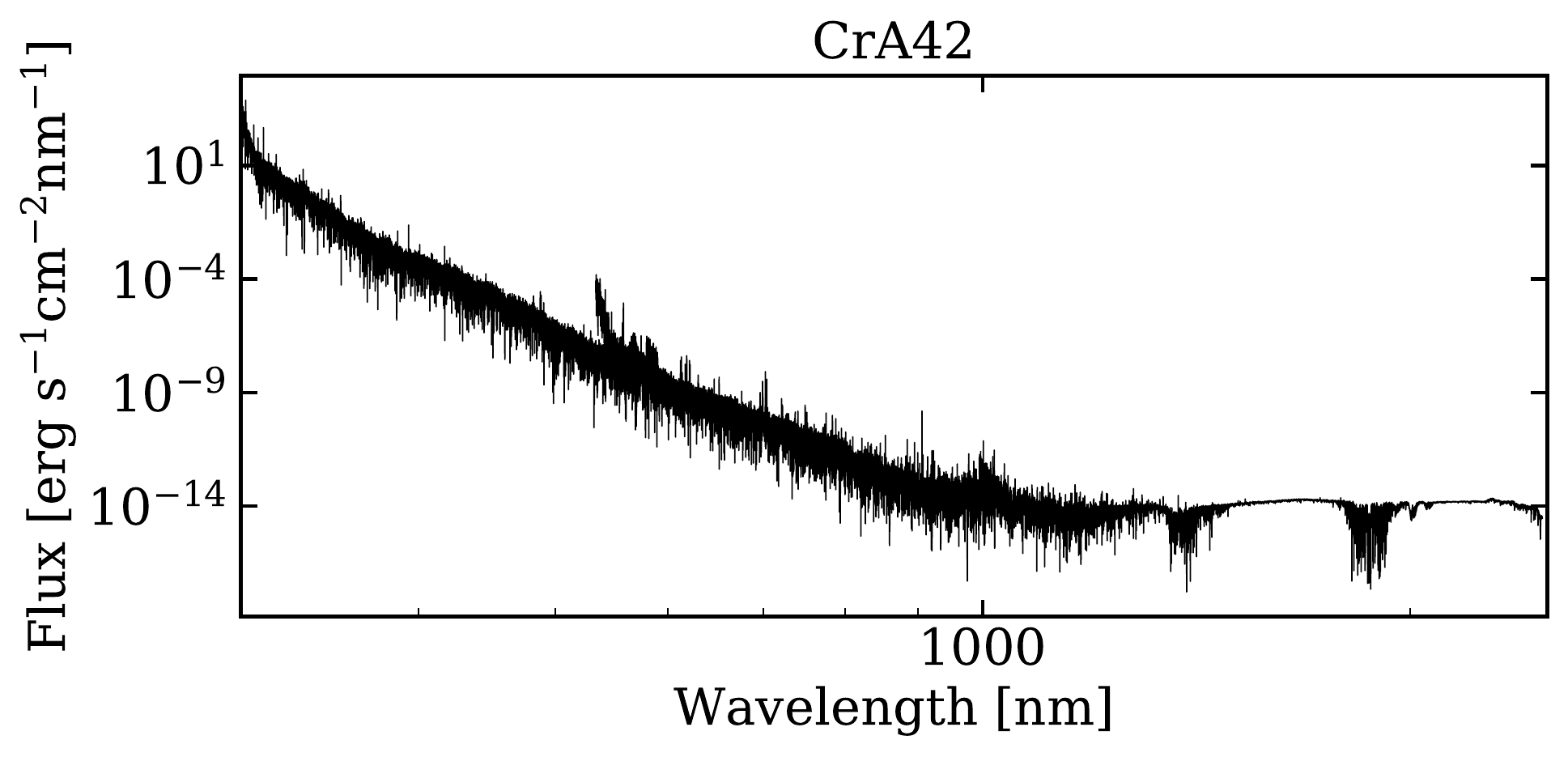}
\phantomcaption
\end{subfigure}%

\begin{subfigure}{.86\linewidth}
  \centering
  \includegraphics[width=\linewidth]{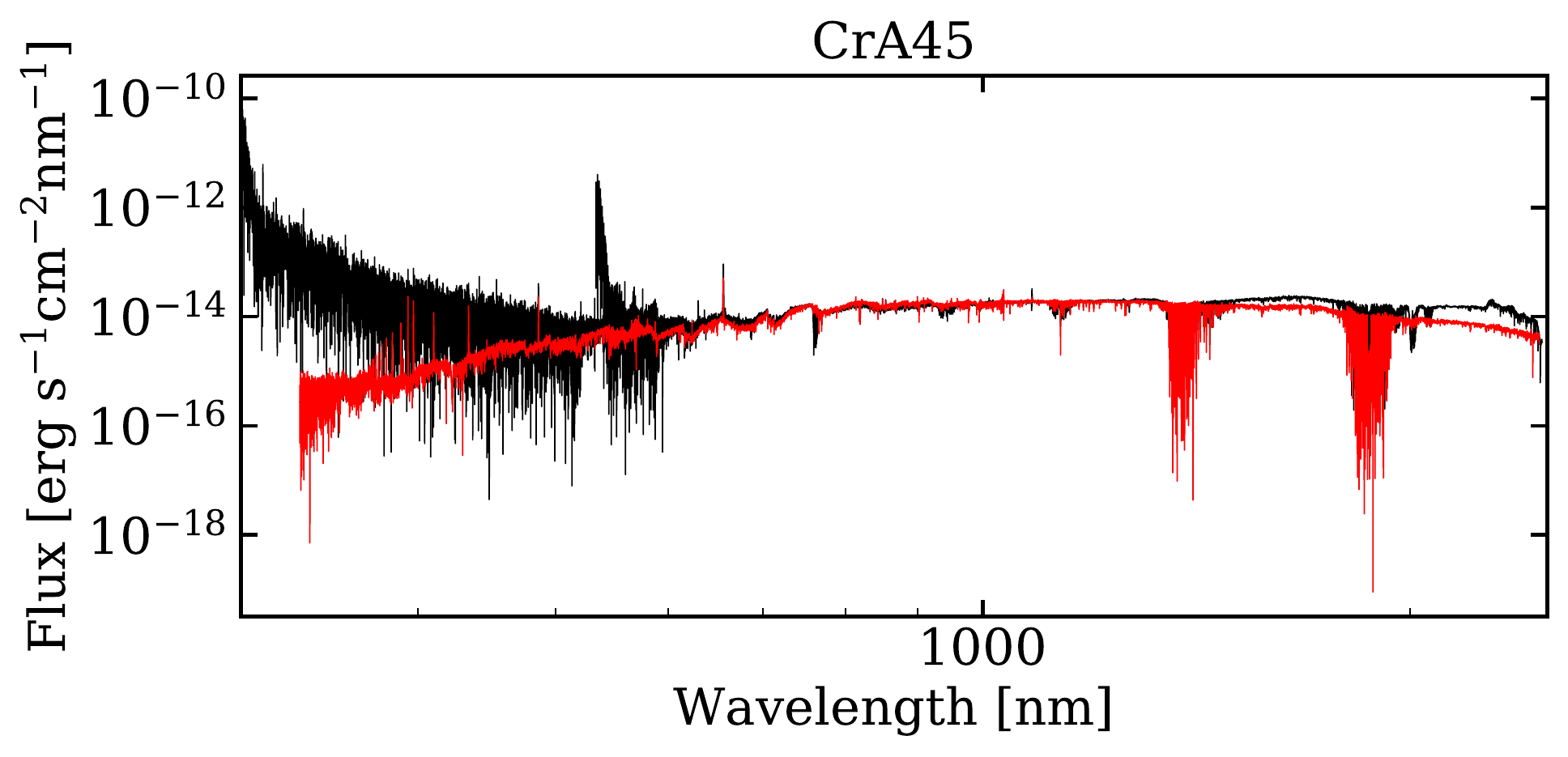}
\phantomcaption
\end{subfigure}

\begin{subfigure}{.86\textwidth}
  \centering
  \includegraphics[width=\linewidth]{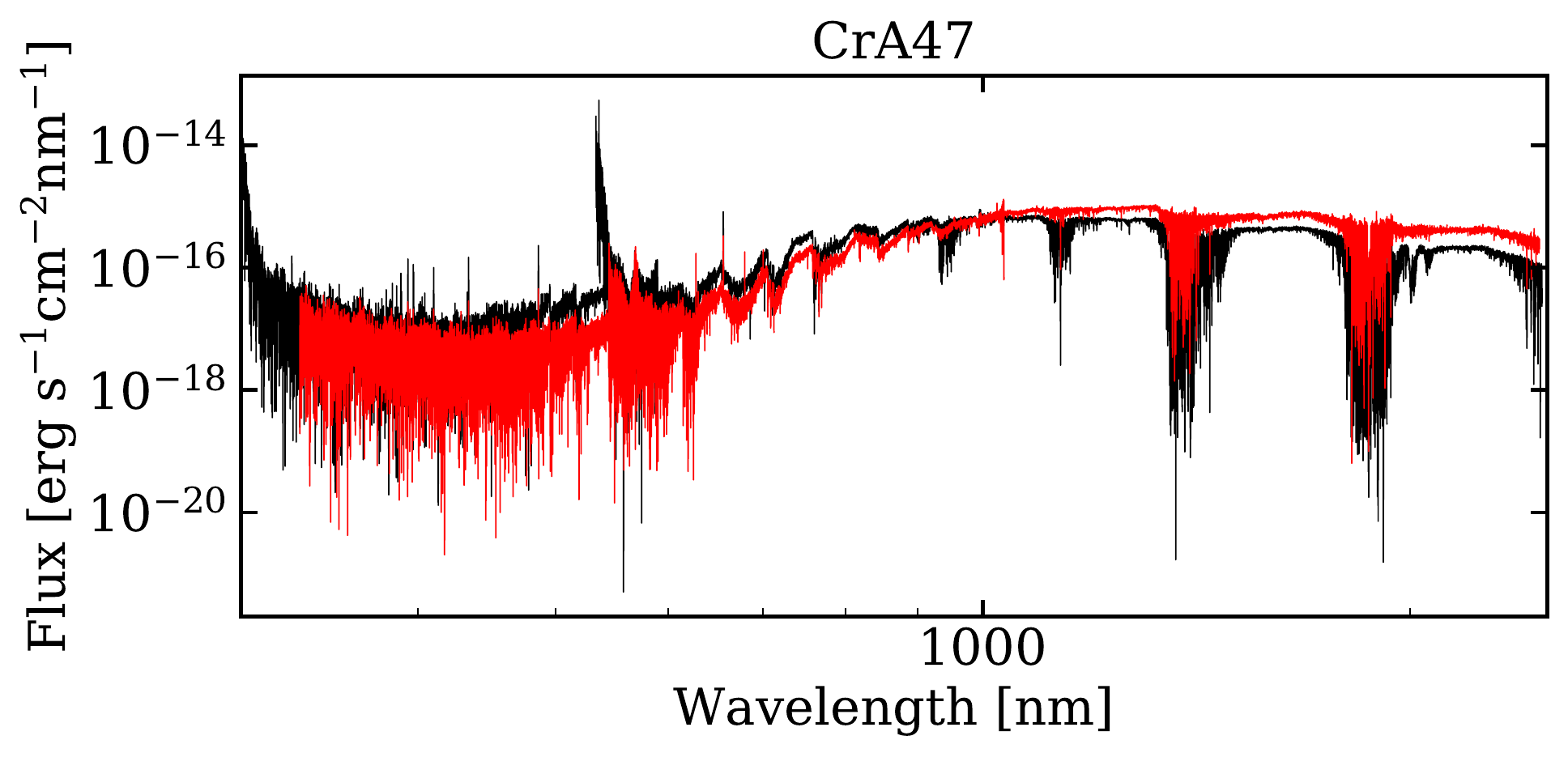}
\phantomcaption
\end{subfigure}
\caption{Continued}
\end{figure*}

\clearpage

\begin{figure*}\ContinuedFloat
\begin{subfigure}{0.86\textwidth}
  \centering
  \includegraphics[width=\linewidth]{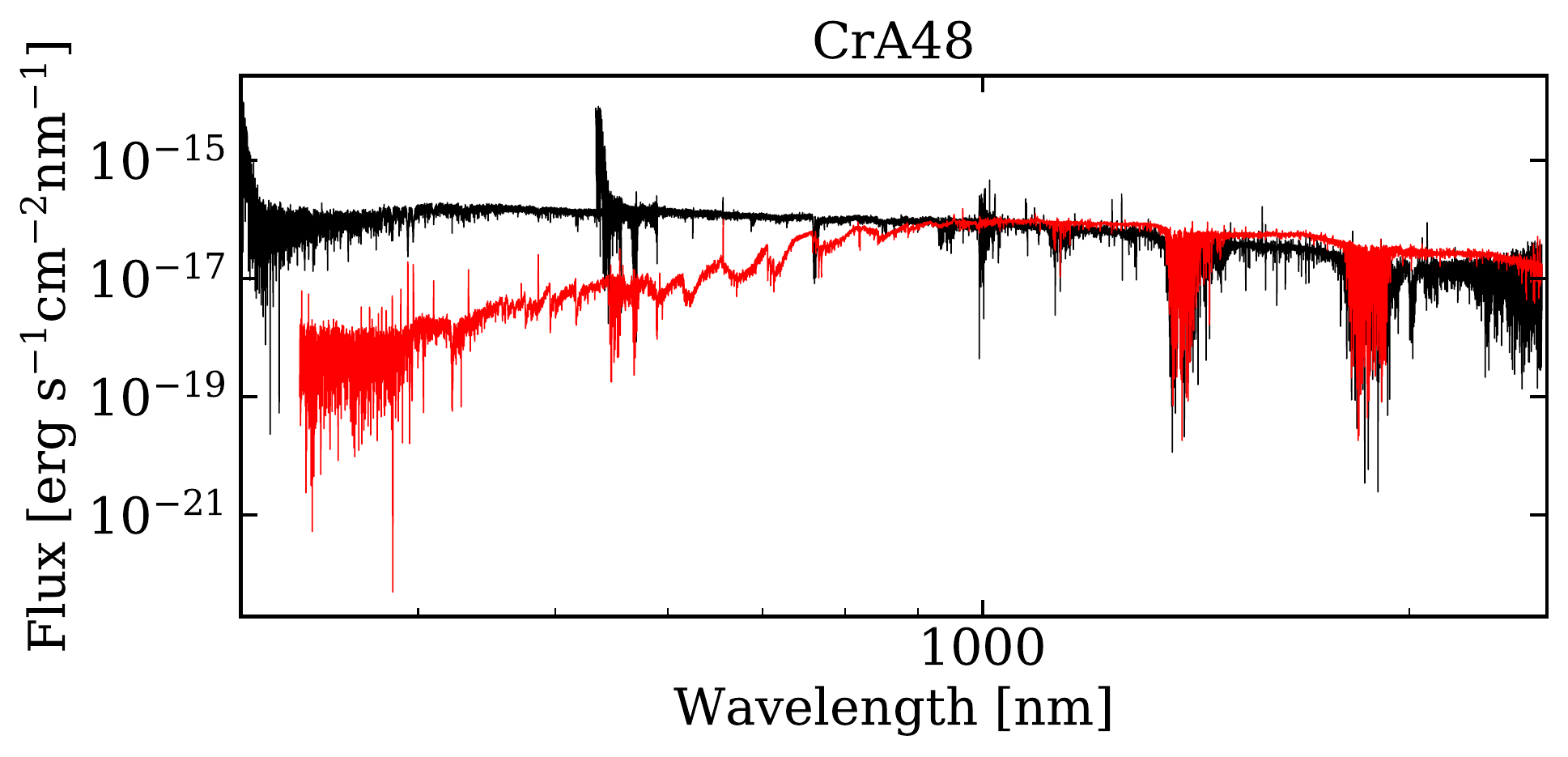}
\phantomcaption
\end{subfigure}%

\begin{subfigure}{.86\linewidth}
  \centering
  \includegraphics[width=\linewidth]{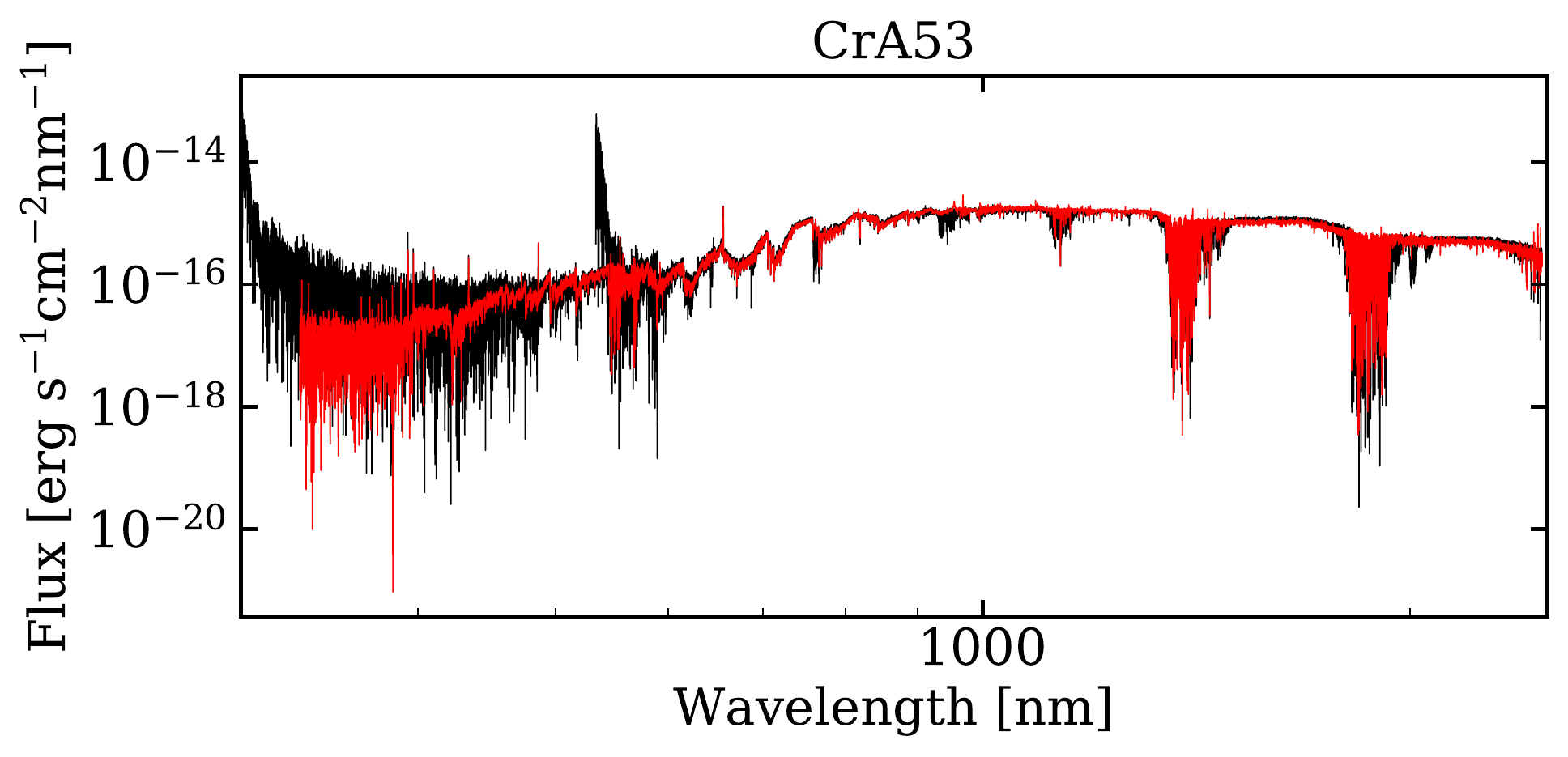}
\phantomcaption
\end{subfigure}

\begin{subfigure}{.86\textwidth}
  \centering
  \includegraphics[width=\linewidth]{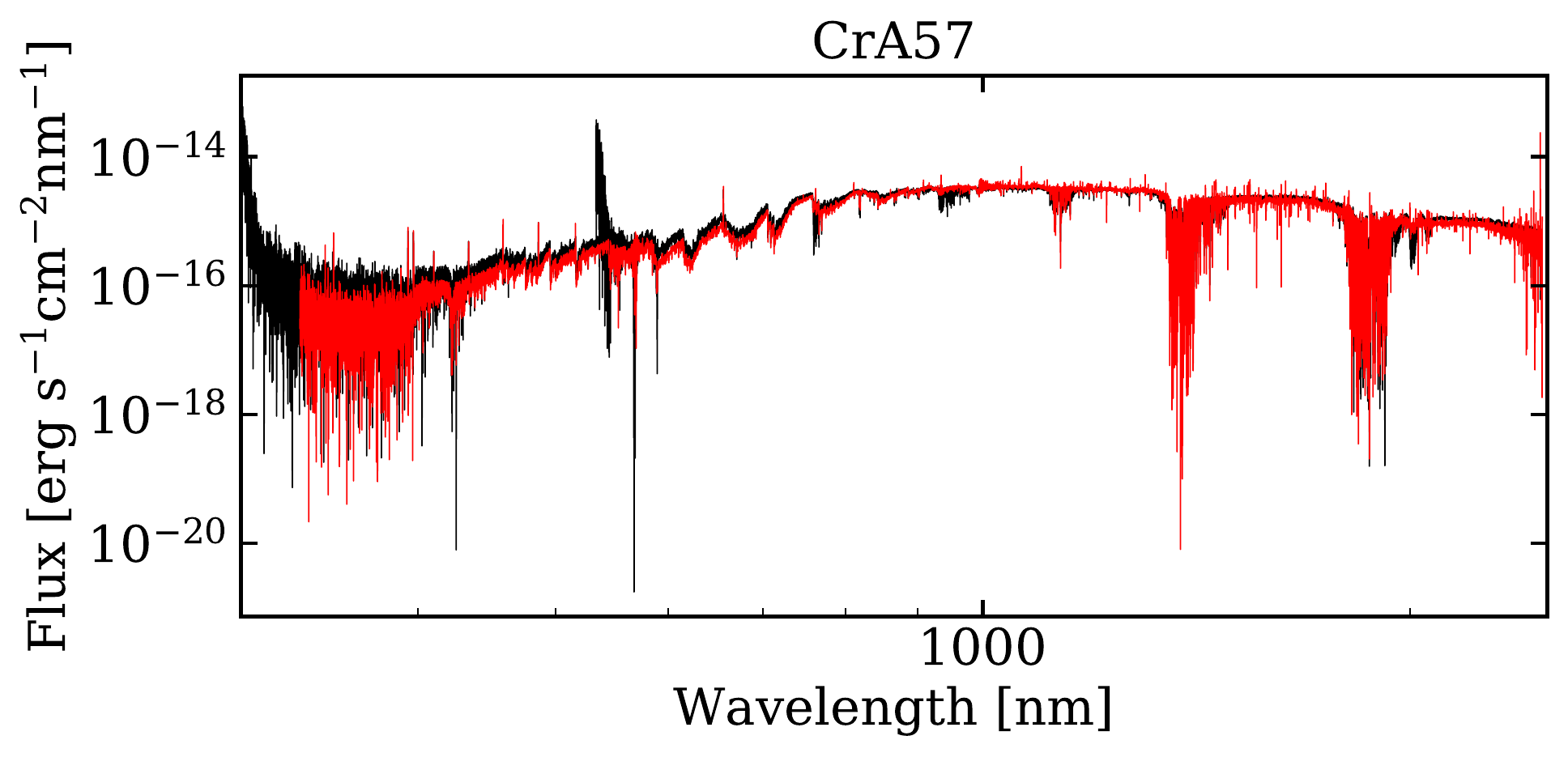}
\phantomcaption
\end{subfigure}
\caption{Continued}
\end{figure*}
\clearpage

\begin{figure*}\ContinuedFloat
\begin{subfigure}{.86\textwidth}
  \centering
  \includegraphics[width=\linewidth]{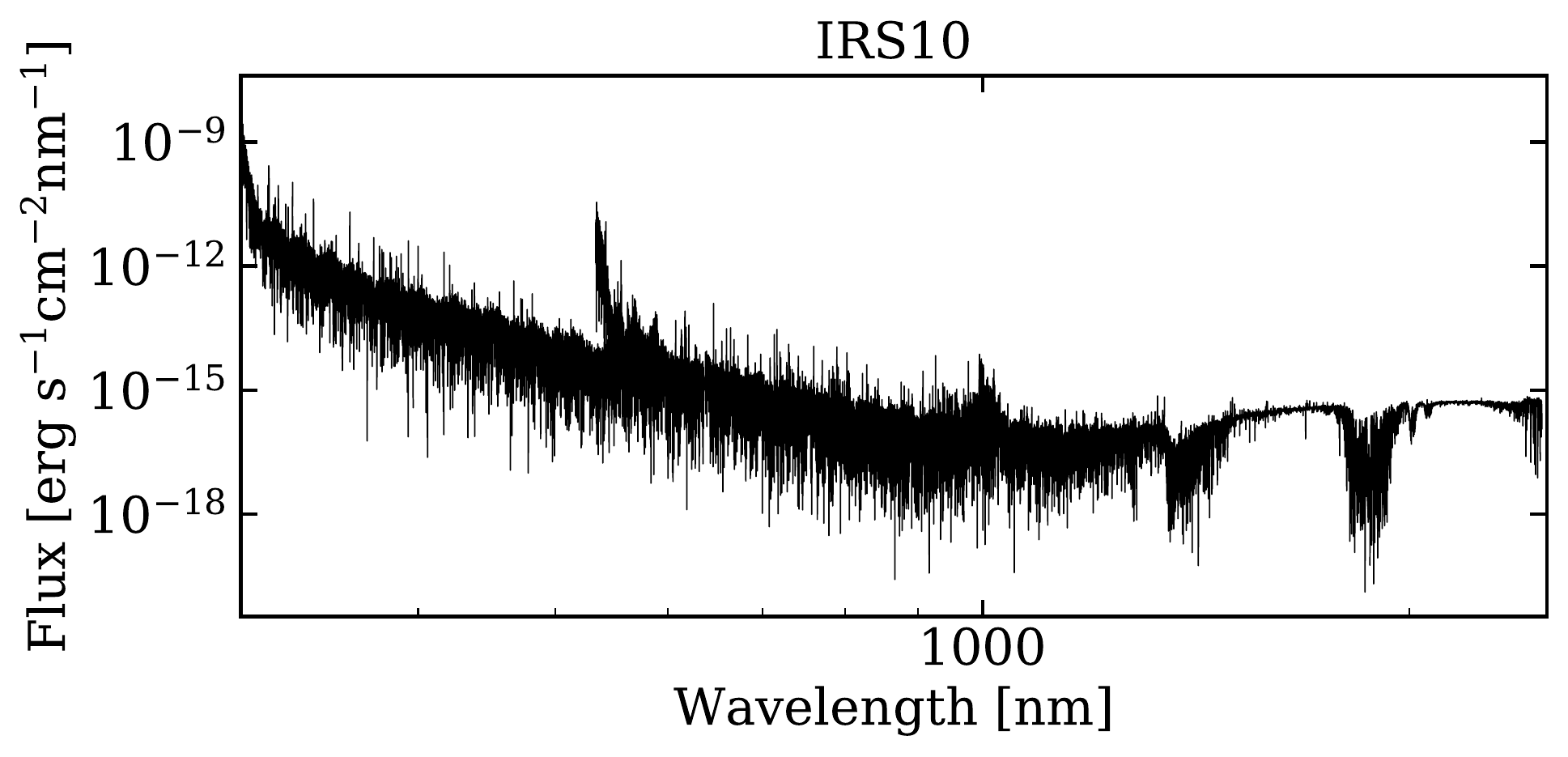}
\phantomcaption
\end{subfigure}
\caption{Continued}
\label{fig:spectra}
\end{figure*}

\end{document}